\documentclass[preprint,floatfix,aps,prd,showpacs,nofootinbib,amsmath,amssymb,amsfonts,superscriptaddress]{revtex4-2}
\usepackage[utf8]{inputenc}
\usepackage{bm}
\usepackage{amsmath,amssymb}
\usepackage{graphicx}
\usepackage{physics}
\usepackage{dsfont}
\usepackage[normalem]{ulem}
\usepackage{hyperref,url}
\usepackage{mathrsfs}
\usepackage{calrsfs}
\usepackage{tikz}
 \usepackage{comment}

\usepackage[caption=false]{subfig}
\renewcommand{\vec}[1]{\mathbf{#1}}
\usepackage[active]{srcltx}

\expandafter\ifx\csname package@font\endcsname\relax\else
\expandafter\expandafter
\expandafter\usepackage
\expandafter\expandafter
\expandafter{\csname package@font\endcsname}%
\fi
\hypersetup{
	colorlinks=true,       
	linkcolor=blue,          
	citecolor=blue,        
}

\usepackage[section]{placeins}

\usepackage{fancybox}
\usepackage{soul}

\begin{document}
\title{Can Q-balls describe cosmological and galactic dark matter?}

\author{Susobhan Mandal}
\email{sm12ms085@gmail.com}

\author{S. Shankaranarayanan}
\email{shanki@iitb.ac.in}

\affiliation{ Department of Physics, 
Indian Institute of Technology Bombay,
Mumbai - 400076, India }

\date{\today}

\begin{abstract}
The Cold Dark Matter (CDM) hypothesis accurately predicts large-scale structure formation and fits the Cosmic Microwave Background temperature fluctuations (CMB). However, observations of the inner regions of dark matter halos and dwarf galaxy satellites have consistently posed challenges to CDM. On the other hand, the Modified Newtonian Dynamics (MOND) hypothesis can explain galactic phenomena but fails to account for the complex shape of the CMB and matter power spectra. CDM and MOND are effective in nearly mutually exclusive regimes, prompting the question: Is there a physical mechanism where CDM and MOND share a common origin? Q-balls, which are localized, non-topological solitons, can be a bridge between the two hypotheses. Q-balls formed in the early Universe can mimic CDM at cosmological scales. Interestingly, Q-balls can exhibit MOND-like behavior in the late Universe at galactic scales, providing a unified framework. Specifically, we demonstrate that millicharged composite Q-balls formed from complex scalar fields, decoupled from the background radiation, can naturally arise during the radiation-dominated epoch. From the matter-radiation equality, we also obtain the mass of Q-balls to be $1~{eV}$, which are much smaller than the electron mass. Using the constraints from the invisible decay mode of ortho-positronium, we obtain $Q < 3.4 \times 10^{-5}$. We also establish an upper bound on the number density of Q-balls, which depends on the charge of the Q-ball and the small initial charge asymmetry. Furthermore, we demonstrate that the MOND naturally emerges at the galactic scale within the framework of our Q-ball model.
\end{abstract}

\maketitle
\newpage

\section{Introduction}
The most convincing evidence for dark matter (DM) is found at the cosmological scales~\cite{liddle1993cold, navarro1996structure, dubinski1991structure, diemand2011structure, silk1993clumpy, klypin2001resolving}. The $\Lambda$-Cold-Dark-Matter ($\Lambda$CDM) model, in which DM is composed of collisionless particles, fits the temperature fluctuations of the Cosmic Microwave Background (CMB), the matter power spectra, and the abundance and mass function of galaxy clusters extraordinarily well. The foundation of the conventional cold dark matter hypothesis is the existence of non-relativistic, collision-free dark matter particles \cite{liddle1993cold, navarro1996structure, dubinski1991structure, diemand2011structure, silk1993clumpy, klypin2001resolving}. Also, dark matter constituents cannot possess an electric charge comparable to that of electrons unless they are extremely heavy~\cite{Bertone:2010zza}. However, this does not rule out the intriguing possibility of electrically millicharged particles~\cite{Prinz:1998ua,deMontigny:2023qft}. Millicharged particles have an electric charge $e' = \varepsilon e$, where $e$ is the electron charge and $\varepsilon \ll 1$. These particles can be either bosons or fermions, and they naturally arise in a wide range of models~\cite{Prinz:1998ua,deMontigny:2023qft}. On cosmological scales, CDM accurately predicts the formation of structure. 

At smaller distance scales, however, the accurate picture is nebulous. As galaxy simulations and measurements have advanced, the CDM paradigm has encountered several challenges. Contradictory predictions concerning structure on galactic and sub-galactic scales, however, seem to result from it \cite{kusenko2001q, spergel2000observational, wandelt2001self}. 
Local group dwarf satellite galaxies present the main obstacles. Dwarf satellites are excellent candidates for in-depth studies of DM microphysics due to their DM dominance. With the discovery of ultra-faint dwarfs~\cite{willman2005new, belokurov2007cats, walsh2008invisibles}, the old "missing satellite" problem~\cite{klypin1999missing, moore1999dark} has been gradually resolved, and other, more pressing problems have emerged. Recent attempts to match populations of simulated subhaloes and observed Milky Way (MW) dwarf galaxies uncovered a ``too big to fail" issue. The most massive black halos are too dense to contain the brightest MW satellites. Even more perplexing is that most MW and Andromeda (M31) satellites co-rotate within immense planar structures. This can not be explained within the $\Lambda$CDM model~\cite{Chae:2023prf}. Also, recently, it has been suggested that the growth rate of perturbations is higher than predicted by the $\Lambda$CDM model~\cite{PhysRevLett.131.111001}.

These issues raise the possibility or implication that dark matter is not cold or collision-free. Weakly interacting massive particles (WIMPS) \cite{petriello2008dama, steigman2012precise, roszkowski2018wimp, schumann2019direct} or axions \cite{duffy2009axions, sikivie2010dark, visinelli2009dark, holman1983axions, marsh2014model, klaer2017dark} the two generally accepted possibilities for dark matter, would be excluded in that case. The defining feature of the thermal WIMP is its relic abundance naturally explained by the freeze-out process with a weak-scale cross-section. This cross-section would account for the elusive non-gravitational interactions of dark matter (DM). WIMP candidates naturally appear around the weak scale in many theories beyond the standard model (BSM). While a simple thermal WIMP is not the only possibility for DM, it is a compelling scenario that demands definitive testing.
If WIMPs constitute the DM in the Universe, including our Galaxy's DM halo, they should be ubiquitous, even in our immediate vicinity. This raises the question of directly detecting these WIMPs in the laboratory. This possibility was first explored by M Goodman and E Witten in 1985. They proposed that WIMPs elastically scattering off nuclei of chosen detector materials might leave recoiling nuclei with detectable kinetic energies.
Goodman and Witten suggested that coherent elastic scattering of WIMPs off nuclei, with the cross-section proportional to $A^{2}$ ($A$ is the mass number of the nucleus), could yield a detectable rate of scattering events. Subsequently, numerous experiments worldwide, employing diverse detection techniques and materials, have sought WIMPs. However, to date, none of these experiments have definitively claimed WIMP detection.

In contrast, Modified Newtonian Dynamics (MOND)~\cite{milgrom1981modification,milgrom1982modification, PhysRevLett.127.161302} provides a radical alternative to DM by modifying the Newtonian force law. According to MOND, the modification to Newtonian force law occurs at low acceleration. Thus, MOND can be understood as either a change to the Poisson equation that modifies gravity or a shift in inertia that modifies inertia by breaking the inertial and gravitational mass equivalence. This empirical force law has astonishingly explained a broad range of galactic events~\cite{Chae:2023prf}. It predicts asymptotically flat rotation curves for spiral galaxies and provides an excellent fit to exact rotation curves~\cite{sanders2002modified}. The only free parameter is the critical acceleration $a_0$, whose best-fit value is the Hubble constant. Interestingly, the Baryonic Tully Fisher Relation (BTFR)~\cite{mcgaugh2000baryonic,mcgaugh2005baryonic} is a direct result of this force law deep within the MOND regime. 

In $\Lambda$CDM, galaxies are surrounded by extensive DM halos, so a merger cannot be avoided~\cite{nipoti2007galaxy}.
In contrast, in MOND, there is only stellar dynamical friction so that a merger can be avoided~\cite{nipoti2007galaxy}. According to MOND, tidal dwarf galaxies should have flat rotation curves and reside on the BTFR, consistent with NGC5291's dwarfs. On extragalactic scales, however, MOND faces more severe problems (see~\cite{angus2011abundance,angus2006proof, sanders2003clusters,aguirre2001problems}). MOND and CDM are, therefore, effective in almost mutually exclusive regimes. The $\Lambda$-CDM model can explain the expansion and linear growth histories and the abundance of clusters, but on a galactic scale, it has certain limitations. MOND explains the observable features of galaxies reasonably well in general, notably the empirical scaling relations. However, it is highly improbable that it can be consistent with the complex shape of the CMB and matter power spectra. 

This naturally raises the question: Is there a framework where the DM behaves as a CDM in the cosmological scales and can mimic MOND in the galaxy scales? In other words, is there a mechanism where CDM and MOND share a common ancestry? To our knowledge, it is difficult to have CDM and MOND behaviors through particle dark matter models. As mentioned earlier, WIMPS or Axions cannot connect these theories in their respective mutually exclusive domains. More than that, WIMPs and axions cannot explain the low-energy theory of phonon modes at the galactic scale as their rest mass energy is very high.
Recently, the interest in Primordial Blackholes (PBHs) as a dark matter candidate over particle dark matter candidates is to fill this gap~\cite{frampton2010primordial, belotsky2014signatures, clesse2018seven, ozsoy2023inflation, niemeyer1999dynamics, garcia2020primordial, carr2021constraints, carr2016primordial, sasaki2018primordial, villanueva2021brief, carr2020primordial}. Also, PBHs are not subject to the Big Bang nucleosynthesis (BBN) constraints of Baryons, making them non-baryonic entities that exhibit similar characteristics to CDM particles. 

This work proposes an alternative mechanism with the same origin but behaves differently in the cosmological and galactic scales. 
We demonstrate that non-topological solitons --- the Q-balls --- can be formed in the early Universe and mimic CDM at the cosmological scales. Specifically, we show that Q-balls from complex scalar field \emph{decoupled} from the background radiation can naturally form in the radiation-dominated epoch. In contrast, they mimic MOND at galactic scales in the late Universe.
Q-balls can naturally form in a wide range of particle physics models \cite{lloyd2022q, kusenko1997small, coleman1985q, cohen1986evaporation, kusenko1998supersymmetric, loiko2018q, anagnostopoulos2001large, safian1988some, tamaki2014large, safian1988some, Kusenko:2005du, brihaye2008interacting} and can be produced in the early Universe, and can be stable. {Recently, the existence of Q-balls in the dark sector of the Universe and its astrophysical consequences have gained interest~\cite{Ansari:2023cay}.}

For a class of complex scalar field theory, we show that Q-ball configurations exist in the recombination epoch during the radiation-dominated (RD) era in the early Universe. We show that the Q-balls formed satisfy the two conditions --- stability and existence. We show that the density perturbations that reenter during the recombination epoch increase the production of Q-balls and compute the number density of the thin-wall Q-balls in the Universe. We also compute the upper bound on the number density of Q-balls based on its global $U(1)$ charge and small primeval charge asymmetry, which might be created due to primordial anomaly due to helical magnetic fields similar to the case shown in Ref.~\cite{Kushwaha:2021csq}. It is also possible to generate charge asymmetry due to the helicity of the gravitational waves~\cite{PhysRevLett.96.081301}. From the matter-radiation equality, we also obtain the mass of Q-balls to be $1~{\rm eV}$, which are much smaller than the electron mass. Combined with the invisible decay mode of ortho-positronium leads to $Q < 3.4 \times 10^{-5}$ \cite{PhysRevD.88.117701, PhysRevD.75.032004, PhysRevD.75.075014}. Suggesting that the millicharged Q-balls are DM candidates responsible for the early structure. Since the Q-balls decay as $1/a^3(\eta)$ in the early epoch, they behave like CDM in the cosmological scales. 

Later, we look at the possibility of forming Bose-Einstein condensate (BEC) and superfluidity by these Q-balls in the present Universe (dominated by dark energy). We find the Q-balls can indeed form these phases of matter, which eventually lead to the law predicted by Modified Newtonian dynamics at the galactic scale~\cite{berezhiani2015theory, khoury2022dark}. Thus, we show that the \emph{cosmological and galactic dark matter} share a common ancestry, representing distinct phases of a single background fluid. 

The possibility that galactic DM can be a superfluid has recently attracted attention~\cite{khoury2022dark, berezhiani2015theory, Berezhiani:2015pia}. This is based on two key ideas: The first notion that the DM forms a superfluid within galaxies with a coherence length equal to the size of the galaxies is widespread \cite{khoury2022dark, berezhiani2015theory}\footnote{In Ref.~\cite{Mistele:2022vhh}, galactic rotation curves are fitted to test the superfluid dark matter model.}. Second, the DM superfluidity phenomenon occurs frequently if the DM particle is sufficiently light and has a significant self-interaction. 
A superfluid is a state of matter where impurity particles flow without dissipation as long as they remain below a critical velocity, a phenomenon known as Landau's criterion~\cite{astrakharchik2004motion}. Superfluidity and Bose-Einstein condensation are intimately related phenomena. In order to acquire a superfluid phase, Bose-Einstein condensation must first occur; however, the converse is not true, as the superfluidity property disappears in the absence of interactions.
Thus, we show that Q-balls can be considered a DM candidate, which acts as a composite object at the cosmological scale and behaves as collective excitation on the galactic scale. 

The present article is structured as follows. In section \ref{section 1}, we consider a complex scalar field theory in the spatially flat FRW spacetime at the RD epoch. For adiabatic evolution, we derive the equations governing the radial profile Q-balls. We include thermal corrections in the RD epoch and show that the thermal corrections enhance the formation of Q-balls. In section \ref{section 6}, we discuss why Q-balls can be considered a dark matter candidate. Further, we compute the rate of formation of Q-balls and their number density in the MD epoch of the Universe. In section \ref{section 7}, we look at the possibility of forming BEC and superfluid phase of matter by Q-balls, which eventually can resolve issues faced by the Modified Newtonian Dynamics and $\Lambda$CDM model simultaneously. Finally, we end this article by concluding our main results in the section \ref{section Discussion}.

\section{Complex scalar field theory in early Universe}\label{section 1}

The general invariant line element describing the expansion of the spatially flat Universe is given by
\begin{equation}\label{eq. 2}
ds^{2} = -dt^{2} + a^{2}(t) d\boldsymbol{x}^{2} = a^{2}(\eta)( - d\eta^{2} + d\boldsymbol{x}^{2}),
\end{equation}
where $t(\eta)$ is the cosmic  (conformal) time, $\boldsymbol{x}$ refers to the 3-spatial coordinates, and $a(t) [a(\eta)]$ is the scale factor in cosmic [conformal] time. On this background, we consider a test complex scalar field theory ($\Phi$) described by the following action:
\begin{equation}\label{eq. 4}
S = - \int d^{4}x \sqrt{-g} \, \Big[g^{\mu\nu}\partial_{\mu}\Phi\partial_{\nu}\Phi^{*} + 
U(\Phi^{*}\Phi)\Big]
\end{equation}
where $\Phi$ is a complex scalar field and $U$ is the field potential. 
In Appendix \eqref{app:U1gauge} we explicitly show that this 
action describes the Q-ball configuration for the FRW background and the interaction with the U(1) gauge field is irrelevant for the FRW background. In conformal time, the above action reduces to:
\begin{equation}\label{eq. 5}
\begin{split}
S & = 
 \int d^{4}x \ \Big[a^{2}{\Phi}' {\Phi'}^{*} - a^{2}\vec{\nabla}\Phi.\vec{\nabla}
 \Phi^{*} - a^{4}U(\Phi^{*}\Phi)\Big].
\end{split}
\end{equation}
where prime denotes derivative \textit{w.r.t} the conformal time. Redefining the field variable in the following manner
\begin{equation}\label{eq. 6}
\phi = a(\eta)\Phi, \ \phi^{*} = a(\eta)\Phi^{*}
\end{equation}
the above action becomes:
\begin{equation}\label{eq. 7}
\begin{split}
S & = \int d^{4}x \ \Big[(\phi' - \mathcal{H}\phi)(\phi'^{*} - \mathcal{H}\phi^{*}) - 
\vec{\nabla}\phi.\vec{\nabla}\phi^{*} - \bar{U}(\phi^{*}\phi)\Big]\\
 & = \int d^{4}x \ \Big[\phi'\phi'^{*} - \mathcal{H}(\phi\phi'^{*} + \phi'\phi^{*}) + 
\mathcal{H}^{2}|\phi|^{2} - \vec{\nabla}\phi.\vec{\nabla}\phi^{*}\\
 & - \bar{U}(\phi^{*}\phi)\Big],
\end{split}
\end{equation}
where 
\begin{equation}\label{eq. 8}
\mathcal{H} \equiv \frac{a'}{a},~~
\bar{U}(\phi\phi^{*}) \equiv  a^{4}U(\Phi\Phi^{*})  \, .
\end{equation}
By doing integration by parts, we may write the action as:
\begin{equation}\label{eq. 9}
S = \int d^{4}x \ \Big[\phi'\phi'^{*} + (\mathcal{H}' + \mathcal{H}^{2})|\phi|^{2} - 
\vec{\nabla}\phi.\vec{\nabla}\phi^{*} - \bar{U}(\phi^{*}\phi)\Big].
\end{equation}
We may note the following:
\begin{equation}\label{eq. 10}
\begin{split}
\mathcal{H}'(\eta) & + \mathcal{H}^{2}(\eta) = \frac{d}{d\eta}\left(\frac{1}{a}\frac{da}{d\eta}
\right) + \left(\frac{1}{a}\frac{da}{d\eta}\right)^{2}\\
 & = a^{2}[\dot{H} + H^{2}] + a^{2}H^{2} = a^{2}(\dot{H} + 2H^{2})
\end{split}
\end{equation}
where `dot' represents derivative \textit{w.r.t} the coordinate time, and
\begin{equation}\label{eq. 11}
\dot{H} + 2H^{2} = H^{2}(1 - q), \ q = - \frac{\ddot{a}a}{\dot{a}^{2}}.
\end{equation}
In the case of de Sitter, $q = - 1$, and for all other power-law expansions, $q > -1$. Interestingly, for the radiation-dominated (RD) epoch $q = 1$ and hence $\mathcal{H}' + \mathcal{H}^{2}$ vanishes. Thus, in the RD epoch, except for the potential term, all the other terms in the above action do not have any explicit time-dependence. Thus, studying the Q-ball formation at any time in this epoch is possible. In the rest of this section, we explicitly show the formation of Q-balls from the charged scalar field ($\Phi$) in this era through the thermal phase transition.

\subsection{Conditions for Q-ball formation during RD epoch}

As a first step, we first obtain the physical conditions that are required for the Q-ball formation during RD epoch.
To go about this, we obtain an associate conserved charge invariant under global $U(1)$ transformation. From Noether's theorem, the expression of the total conserved global $U(1)$ charge is:
\begin{equation}\label{eq. 12}
\mathcal{Q} = - i\int d^{3}x (\phi'^{*}\phi - \phi'\phi^{*}).
\end{equation}
Further, the Hamiltonian corresponding to action \eqref{eq. 9}:
\begin{equation}\label{eq. 13}
H_{\phi} = \int d^{3}x \Big[\phi'\phi'^{*} + 
\vec{\nabla}\phi.\vec{\nabla}\phi^{*}  + \bar{U}(\phi^{*}\phi)\Big].
\end{equation}
Now, we define the following quantity in which we are interested in 
\begin{equation}\label{eq. 14}
\begin{split}
\mathcal{E}_{\lambda} & = H_{\phi} + \lambda (Q - \mathcal{Q}) = \int \!\! d^{3}x \Big[|\phi' - i\lambda \phi|^{2} + \vec{\nabla}\phi.\vec{\nabla}\phi^{*}\\
& + \bar{U}(\phi^{*}\phi) - \lambda^{2}|\phi|^{2}\Big]  + \lambda Q,
\end{split}
\end{equation}
since we want to obtain a field configuration by minimizing the energy subjected to the global $U(1)$ charge conservation, which acts as a constraint, and $\lambda$ is the Lagrange multiplier. To minimize the above quantity, we need to define 3-space coordinates explicitly. We consider $d\boldsymbol{x}^{2} = dr^2 + r^2 (d\theta^2 + \sin^2\theta d\varphi^2)$, where $r$ is the radial coordinate and $(\theta, \varphi)$ are the angular coordinates.

As mentioned above, the action \eqref{eq. 9} has no explicit time-dependence (except the potential term). Hence, at a given instant of time $\eta = \eta_{0}$, we take the following ansatz for the field:
\begin{equation}\label{eq. 15}
\phi(\eta_{0},r) = \frac{1}{\sqrt{2}}\sigma(r, \eta_{0}) e^{i\lambda\eta_{0}} \, .
\end{equation}
Note that the $l = 0$ spherical mode will cost the least energy for the system. If a Q-ball does not form in this configuration, then Q-balls in another configuration are highly unlikely, and hence, we only consider $l = 0$ mode. Substituting the above expression in Eq.~(\ref{eq. 14}) leads to:
\begin{equation}\label{eq. 16}
\mathcal{E}_{\lambda} = 4 \pi \int \!\! dr \, r^2 \Big[\frac{1}{2} (\partial_r \sigma)^2 + \bar{U}(\sigma) - \frac{1}{2}\lambda^{2}\sigma^{2}\Big] + \lambda Q.
\end{equation}
In order to evaluate this, we need to assume a form of potential. The condition for an attractive force in the formation of Q-balls cannot be satisfied by a renormalizable single-field potential that is bounded from below \cite{Coleman:1985ki}. Instead, it is necessary to consider multi-field potentials \cite{Friedberg:1976me} or higher-dimensional operators that arise from integrating out heavier fields \cite{Lee:1991ax}. Neglecting quantum corrections, the latter approach leads to polynomial potentials in the scalar field $\sigma$. In our work, we limit ourselves to polynomials with three terms of the form \cite{Heeck:2022iky, Heeck:2020bau}
\[
\bar{U}(\sigma) = m_{\sigma}^{2}\sigma^{2} - \zeta\sigma^{p} + \xi\sigma^{q},
\]
as this represents the minimal form that meets the requirements for the formation of large 
Q-balls. Although the majority of our mathematical analysis holds for arbitrary integer or 
even real exponents satisfying $2 < p < q$, it is important to note that for potentials 
derived within effective field theory, both $p$ and $q$ must be even integers. As a result,
the first obvious choice would be $p = 4, \ q = 6$. It should be emphasized that by retaining 
only the renormalizable terms, one cannot obtain stable Q-balls in a single-field potential. Consequently, non-renormalizable terms must be incorporated into the effective scalar potential. According to standard effective field theory expectations, the $\sigma^{6}$ term, which may 
be generated by the inclusion of additional heavy scalar fields, is anticipated to represent 
the dominant contribution of these non-renormalizable effects. Higher-order terms would then 
be suppressed by additional powers of the mass scale of the heavy scalar fields. Hence, we 
consider the following form:
\begin{equation}\label{eq. 17}
\bar{U}(\sigma) = \frac{1}{2}m_{\sigma}^{2}a^{2}\sigma^{2} - \frac{\zeta}{4}\sigma^{4} + \frac{\xi}{6a^{2}}\sigma^{6},
\end{equation}
where $m_\sigma$ is mass of the field $\sigma$ and we make the appropriate field redefinition mentioned earlier in equation \eqref{eq. 6}. The above potential allows the possibility of a negative quartic coupling while higher dimensional operators restore the stability of the potential. Such non-renormalizable operator to the standard model potential can potentially induce a strong first-order phase transition sufficient to drive baryogenesis~\cite{2005-Grojean.etal-PRD}. Substituting this in Eq.~(\ref{eq. 16}) leads to:
$$
\begin{aligned}
\mathcal{E}_{\lambda} & = 4\pi \int \!\! dr \,  r^2\Big[\frac{1}{2} (\partial_r \sigma)^2 + \frac{1}{2}m_{\sigma}^{2}a^{2}\sigma^{2} - \frac{\zeta}{4}\sigma^{4} + \frac{\xi}{6a^{2}}\sigma^{6}\\
 & - \frac{1}{2}\lambda^{2}\sigma^{2}\Big] + \lambda Q.
\end{aligned}
$$
Minimizing the above quantity \textit{w.r.t} $\lambda$ leads to 
\begin{equation}\label{eq. 19}
\lambda = \frac{Q}{4 \pi \int \!\! dr \, r^2 \ \sigma^{2}}.
\end{equation}
We may note that the Hamiltonian explicitly depends on the scale factor for the field ansatz mentioned above. Considering the scale factor changes adiabatically, we may consider this Hamiltonian an instantaneous Hamiltonian. Substituting the ansatz \eqref{eq. 15} in action \eqref{eq. 9} for the RD epoch, we obtain the following equation of motion for the $\sigma$ field:
\begin{equation}\label{eq. 20}
\frac{d^{2}\sigma}{dr^{2}} + \frac{2}{r}\frac{d\sigma}{dr} = (m_{\sigma}^{2}a^{2} - \lambda^{2})\sigma - \zeta\sigma^{3} + \frac{\xi}{a^{2}}\sigma^{5}
\end{equation}
Defining the dimensionless field variable $\chi = \sigma/m_{\sigma}$ and dimensionless radius $\bar{r}= rm_{\sigma}$, the above equation reduces to the following form
\begin{equation}\label{eq. 21}
\frac{d^{2}\chi}{d\bar{r}^{2}} + \frac{2}{\bar{r}}\frac{d\chi}{d\bar{r}} = (a^{2} - \bar{\lambda}^{2})\chi - \zeta\chi^{3} + \frac{\bar{\xi}}{a^{2}}\chi^{5},
\end{equation}
where
\begin{equation}\label{eq. 22}
\bar{\lambda} = \frac{\lambda}{m_{\sigma}}, \ \bar{\xi} = \xi m_{\sigma}^{2}.
\end{equation}
The equation (\ref{eq. 21}) describes the radial profile of Q-ball for a given instant of conformal time at which the scale factor is fixed. This is one of the key expressions regarding which we want to discuss the following points: First, the mass and the Hexic potential terms explicitly depend on the epoch of expansion. 
Second, while the Hexic term is positive definite at all times, the mass term is not positive definite. In other words, the mass term can be positive or negative depending on the scale factor at that instant. 
Third, even if we consider the time-dependence of $\lambda$, since the field equations are linear, for a range of $\lambda$, the mass term can be positive or negative. This provides the stability condition for the Q-balls: 
\begin{equation}
a > \bar{\lambda}~~\Longrightarrow~~m_{\sigma}a > \lambda    
\label{eq:QBall-Stab}
\end{equation}
The above condition is not sufficient condition for the existence of Q-ball. In order to ensure the existence of the Q-balls in the RD epoch, the field potential must have to satisfy the following condition~\cite{PaccettiCorreia:2001wtt}:
\begin{equation}
\lambda^{2} \geq \text{min}\left(\frac{2\bar{U}}{\sigma^{2}}\right).
\end{equation}
We may note the following
\begin{equation}
0 = \frac{d}{d\sigma}\left(\frac{2\bar{U}}{\sigma^{2}}\right) \implies \sigma = 0, \sigma_{\pm} = 
\pm\sqrt{\frac{3a^{2}\zeta}{\xi}},
\end{equation}
where the non-zero value of $\sigma^{2}$ leads to two minima which we can obtain from the following
relation
\begin{equation}
\frac{d^{2}}{d\sigma^{2}}\left(\frac{2\bar{U}}{\sigma^{2}}\right)\Big|_{\sigma_{\pm}} = 2\zeta > 0.
\end{equation}
As a result, we obtain the following relation
\begin{equation}\label{existence condition 2}
\lambda^{2} \geq \text{min}\left(\frac{2\bar{U}}{\sigma^{2}}\right) = \frac{2\bar{U}}{\sigma^{2}}
\Big|_{\sigma_{\pm}} = a^{2}\left(m_{\sigma}^{2} - \frac{3\zeta^{2}}{16\xi}\right).
\end{equation}
Combining the above two inequalities, we may now say that the existence of Q-balls in RD epoch 
is guaranteed once the condition:
\begin{equation}\label{eq. 25 revised draft}
a^{2}\left(m_{\sigma}^{2} - \frac{3\zeta^{2}}{16\xi}\right) 
< \lambda^{2} < (m_{\sigma}a)^{2}
\end{equation}
is satisfied. Moreover, $\lambda^{2} > 0$ condition implies that $m_{\sigma}^{2} > \frac{3\zeta^{2}}{16\xi}$ must be satisfied for Q-balls to exist. The above inequality leads to
\begin{equation}
\zeta^{2} \leq \frac{16 \bar{\xi}}{3} \, ,
\label{eq:constraint01}
\end{equation}
The above condition restricts the domain of the quartic coupling $\zeta$ in terms of the mass $m_{\sigma}$ and the hexic coupling $\xi$. Moreover, the value of $m_{\sigma}$ is later constrained to be $10$KeV. In contrast, the value of $\xi$ can be constrained from the low-energy behavior of phonon modes (specifically from the dispersion) about the superfluid dark matter phase at the galactic scale. In this manner, one can put constraints on the parameters of the field potential of the Q-ball matter.

To our knowledge, this is the first time the conditions for Q-ball formation have been explicitly obtained in the RD epoch. 
Lastly, as shown in the following subsection, the above condition ensures that Q-balls are formed during the RD epoch. However, we still have to obtain $\sigma(r)$ to determine the size of the Q-balls formed and, hence, obtain $\lambda$. Note that the value of $Q$ is an input parameter \cite{tsumagari2009physics}. We determine the above values within the thin-wall approximation in Appendix  \ref{app:thinwall}. Before we proceed, we show that the thermal field theory corrections enhance the Q-ball formation in RD-epoch.

\subsection{Q-ball formation in early Universe through phase transition}\label{section 5}

Having established a condition for the Q-ball formation in the RD epoch, we need to include the thermal effects on the complex scalar field potential $U(\Phi^*\Phi)$. More specifically, we consider the thermal field theory effects on the Hexic potential \eqref{eq. 17}:
\begin{equation}
U(\sigma) = \frac{1}{2}(m_{\sigma}^{2}a^{2} - \lambda^{2})\sigma^{2}  + \frac{\zeta}{4}\sigma^{4} + \frac{\xi}{6a^{2}}\sigma^{6},
\label{eq:Poten}
\end{equation} 
where $\zeta < 0$. As mentioned earlier, the above potential allows the possibility of a negative quartic coupling while higher dimensional operators restore the stability of the potential. Such non-renormalizable operator to the standard model potential can induce a strong first-order phase transition sufficient to drive baryogenesis~\cite{2005-Grojean.etal-PRD}. For discussion on the near-criticality of Higgs Bosons in which the quartic coupling term becomes negative, see \cite{buttazzo2013investigating, espinosa2015cosmological}.

As we show, due to the finite-temperature corrections~\cite{Brandenberger:1984cz, Boyanovsky:1994pa, Quiros:1999jp}, the quartic term can effectively become more negative, leading to a domain in field space in which the potential becomes attractive. This is not exactly a spontaneous symmetry breaking. However, temperature-related corrections change the effective potential so that the quadratic term remains positive, whereas the quartic term becomes more negative~\cite{Brandenberger:1984cz, Boyanovsky:1994pa, Quiros:1999jp}. This leads to the formation of composite Q-balls. 

\begin{figure}
\includegraphics[height = 7cm, width = 8cm]{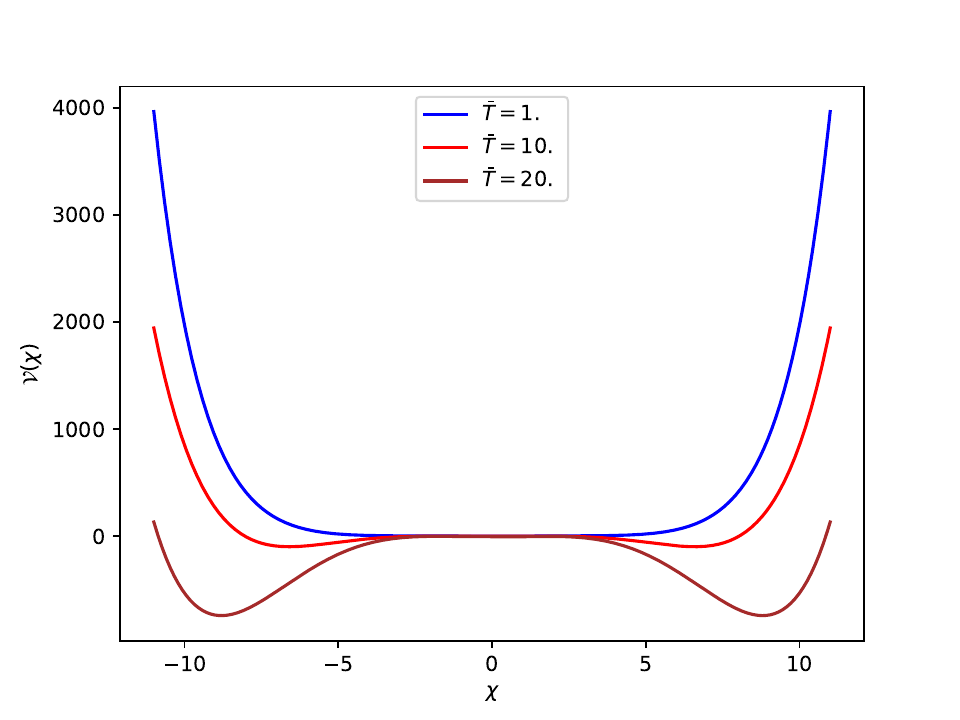}
\includegraphics[height = 7cm, width = 8cm]{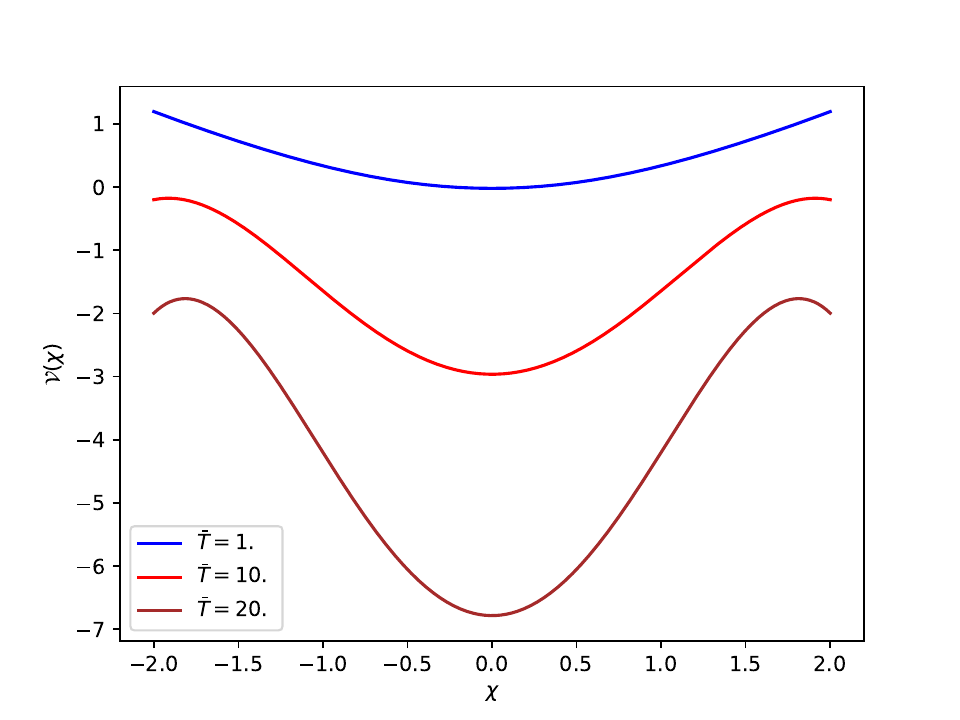}
\caption{Plot of effective field potential for three different temperatures ($\bar{T} = 1, 10, 20$) by setting for $\zeta = - 0.1, \bar{\xi} = 0.01, \ \bar{\lambda} = 0.5, \ a = 1$. The lower plot is a zoomed version of the upper plot. The lower plot shows the enlargement of the magnitude of $\zeta$ (negative) at high temperatures.}
\label{FIG.1}
\end{figure}

Taking into account the finite temperature corrections at the temperature $T$, the effective field potential becomes~\cite{Brandenberger:1984cz,Boyanovsky:1994pa,Quiros:1999jp}
\begin{equation}
\mathcal{V}_{\text{eff}}(\sigma) = \frac{1}{2}(m_{\sigma}^{2}a^{2} - \lambda^{2})\sigma^{2}  + \frac{\zeta}{4}\sigma^{4} + \frac{\xi}{6a^{2}}\sigma^{6} + \frac{1}{2}\text{Tr}\log[\mathcal{G}[\sigma]],
\end{equation}
where
\begin{equation}
\begin{split}
\frac{1}{2}\text{Tr}\log[\mathcal{G}[\sigma]] & = - \frac{TV}{2}\sum_{n}\int\frac{d^{3}k}{(2\pi)^{3}}\log\left(\frac{\mathcal{G}_{0}^{-1}(k,\sigma)}{T^{2}}\right)\\
 & + \frac{TV}{2}\sum_{n}\int\frac{d^{3}k}{(2\pi)^{3}}\log\left(\frac{\mathcal{G}_{0}^{-1}(k)}
{T^{2}}\right),
\end{split} 
\end{equation}
and $\mathcal{G}_{0}^{-1}(k,\sigma) = k^{2} + \omega_{n}^{2} + m_{\sigma}^{2}a^{2} - \lambda^{2} + 3\zeta\sigma^{2} + 5(\xi/a^{2})\sigma^{4}$ where $\omega_{n}$ is the bosonic Matsubara frequency. Carrying out the summation and integration, we obtain the following result at high-temperature expansion
\begin{equation}
\begin{split}
\mathcal{V}_{\text{eff}}(\sigma) & = \frac{1}{2}(m_{\sigma}^{2}a^{2} - \lambda^{2})\sigma^{2} + \frac{\zeta}{4}\sigma^{4} + \frac{\xi}{6a^{2}}\sigma^{6}\\
 & - \frac{1}{24}M_{\sigma}^{2}T^{2} + \frac{1}{12\pi}M_{\sigma}^{3}T\\
 & + \frac{M_{\sigma}^{4}}{64\pi^{2}}\Big[2\gamma_{E} - \frac{3}{2} + \log\left(\frac{M_{\sigma}
^{2}}{16\pi^{2}T^{2}}\right)\Big],
\end{split}
\end{equation}
where $\gamma_{E} = 0.578$ is the Euler-Mascheroni constant and
\begin{equation}
M_{\sigma}^{2} = m_{\sigma}^{2}a^{2} - \lambda^{2} + 3\zeta\sigma^{2} + 5\frac{\xi}{a^{2}}\sigma^{4}.
\end{equation}
Defining $\Delta = m_{\sigma}^{2} a^2 - \lambda^{2}$, we may express the effective field potential
as
\begin{equation}
\begin{split}
\frac{\mathcal{V}_{\text{eff}}(\sigma)}{m_{\sigma}^{4}} & = \mathcal{V}_{\text{eff}}(\chi) = \frac{1}{2}\bar{\Delta}^{2}\chi^{2} + \frac{\zeta}{4}\chi^{4} + \frac{\bar{\xi}}{6a^{2}}\chi^{6}
\\
 & - \frac{1}{24}\bar{M}_{\chi}^{2}\bar{T}^{2} + \frac{1}{12\pi}\bar{M}_{\chi}^{3}\bar{T}\\
 & + \frac{\bar{M}_{\chi}^{4}}{64\pi^{2}}\Big[2\gamma_{E} - \frac{3}{2} + \log\left(
 \frac{\bar{M}_{\chi}^{2}}{16\pi^{2}\bar{T}^{2}}\right)\Big], 
\end{split}
\end{equation}
where $\bar{\xi}$ and $\bar{\lambda}$ are defined in \eqref{eq. 22}, $\chi = \sigma/m_{\sigma}$, 
\begin{equation}
\bar{M}_{\chi}^{2} = a^{2} - \bar{\lambda}^{2} + 3\zeta\chi^{2} + 5\frac{\bar{\xi}}{a^{2}}\chi^{4},
\end{equation}
and 
\begin{equation}
\bar{T} = \frac{T}{m_{\sigma}}, \ \bar{\Delta}^{2} = a^{2} - \bar{\lambda}^{2}.
\label{def:Tbar-Deltabar}
\end{equation}
In Fig. \ref{FIG.1}, we plotted the effective potential for $\zeta = - 0.1, \bar{\xi} = 0.01, \ \bar{\lambda} = 0.5, \ a = 1$ and for three different temperatures $\bar{T} = 1, 10, 20$.
\begin{figure}
\begin{center}
\includegraphics[height = 7cm, width = 9cm]{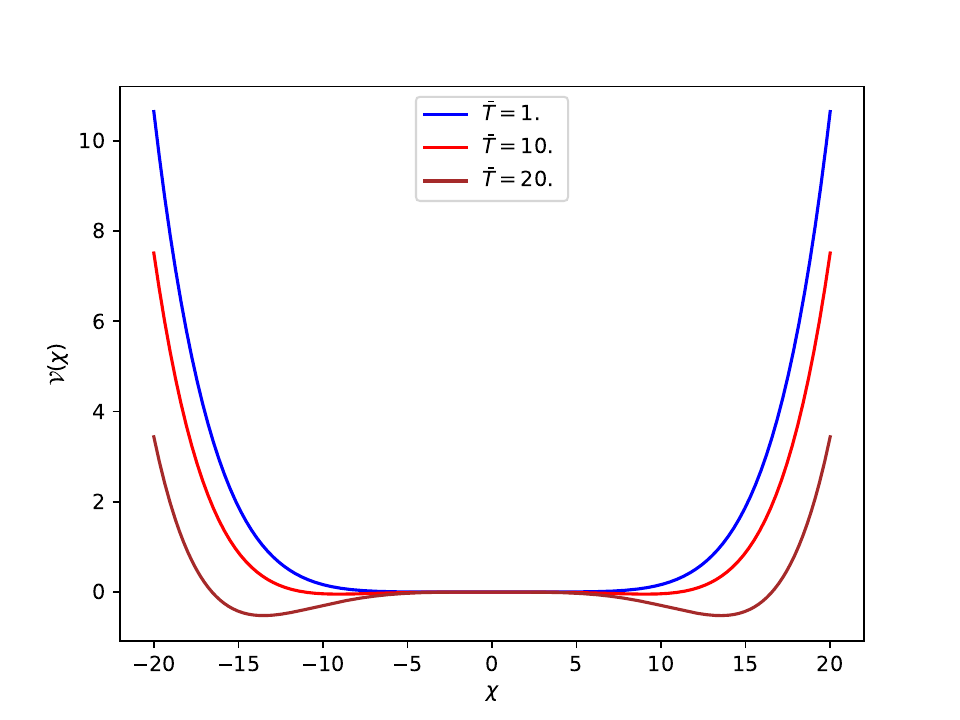}
\end{center}
\caption{plot of effective field potential for three different (scaled) temperatures ($\bar{T} = 1, 10, 20$) by setting for $\zeta = - 10^{-9}, \bar{\xi} = 10^{-6}, \ \bar{\lambda} = 5\times 10^{-5}, \ a = 10^{-4}$.}
\label{FIG.2}
\end{figure} 
In Fig. \ref{FIG.2}, we plotted the same effective potential for the same three scaled temperatures ($\bar{T} = 1, 10, 20$) for different values of the coupling parameters --- $\zeta = - 10^{-9}, \bar{\xi} = 10^{-6}, \ \bar{\lambda} = 5\times 10^{-5}, \ a = 10^{-4}$. Thus, as long as the condition \eqref{eq:QBall-Stab} is satisfied, varied ranges of the coupling parameters lead to composite Q-ball formation.

From the above plots, it is clear that composite Q-balls can form in the early Universe for a broad range of parameters of $\zeta, \bar{\xi}, \bar{\lambda}$ and $\bar{T}$. Since these parameters are rescaled w.r.t $m_{\sigma}$, the mass of the complex scalar field is crucial in determining the composite Q-ball formation in the early Universe. Note that $m_{\sigma}$ is the mass of the scalar field and not the mass of the composite Q-ball.
As shown below, $m_{\sigma}$ can be fixed based on the physical inputs. 

The first input is that if composite Q-balls are dark matter and dominate during the matter-dominated epoch, the Q-balls need to be formed in the RD epoch. In the RD epoch, the free $\Phi$ particles are in equilibrium with the photons, i.e.,
\begin{equation}\label{eq. 59}
\Phi + \Phi^{*}\leftrightarrow 2\gamma \, .
\end{equation}
This can be seen if we consider the action \eqref{eq:U1action} and do a expansion of the form $\Phi = \bar{\Phi} + \delta\Phi$ and consider the ansatzs in \eqref{eq. 6} and \eqref{eq. 15} together and gauged the action defined over the field $\delta\Phi$ and $\delta\Phi^{*}$. This would describe the interactions between elementary $\Phi$-particles with photons and Q-balls. Moreover, given the ansatz for the field $\bar{\Phi}$, the tree-level action reduces effectively to a time-independent real scalar field which do not couple with the gauge field. For more details, see Appendix \ref{app:U1gauge}.

Hence, the formation of Q-balls in the RD epoch requires a primordial $U(1)$ charge asymmetry of the complex scalar field $\Phi$. The primordial $U(1)$ charge asymmetry of the complex scalar field $\Phi$ can be characterized by the conserved quantity~\cite{dolgov1993electric, postma2002solitosynthesis}
\begin{equation}\label{eq. 58}
\eta_{Q} = \frac{n_{\Phi} - n_{\Phi^{*}}}{n_{\gamma}},
\end{equation}
where $n_{\gamma} = 2.4T_{\gamma}^{3}/\pi^{2}$ where $T_{\gamma}$ is the temperature of the photons. As shown recently, the charge asymmetry~\cite{2021-Kushwaha.Shanki-PRD} naturally occurs due to helical magnetic fields generated during inflation~\cite{2020-Kushwaha.Shanki-PRD}. 

The above equilibrium condition implies that the chemical potential for the creation of the free $\Phi$ and $\Phi^*$ particles satisfy the relation $\mu_{\Phi} = - \mu_{\Phi^{*}}$. Thus, the number density of $\Phi$ and $\Phi^*$ free particles are given by:
$$
n_{\Phi^*}=n_{\Phi} e^{-2 \mu_{\Phi} / T} .
$$
Substituting the above expression in Eq. \eqref{eq. 58}, we have: 
\begin{equation}\label{eq. 61}
\eta_{Q} = \frac{n_{\Phi}}{n_{\gamma}}(1 - e^{-2\mu_{\Phi}/T}).
\end{equation}
In the epoch when $T > \mu_{\Phi}$, we have: 
\[
\eta_{Q} \sim \frac{2 n_{\Phi} \mu_{\Phi} }{n_{\gamma} T}
\]
while in the epoch when $T < \mu_{\Phi}$, we have
\begin{equation}\label{eq. 62}
\eta_{Q} = \frac{n_{\Phi}}{n_{\gamma}} \, .
\end{equation}
The second input is the critical temperature $\bar{T}_c \sim 1$ for the Q-ball formation. 
The third input is that the annihilation process between the charged particles $\Phi$ and their anti-particles $\Phi^{*}$, which forms photons, should be sub-dominant compared to the Q-ball formation as Q-ball is a low-energy configuration. 

The fourth input is that once the Q-ball has formed, it needs a robust environment to grow. The process of Q-ball formation through the gradual charge accumulation is efficient as the Q-balls satisfy the condition $E_{Q} = \lambda_{0}Q$ where $\lambda_{0} < m_{\sigma}$ (see the discussion in Appendix \eqref{app:thinwall}) which means that Q-ball is the low-energy configuration compared to free charge particle configuration \cite{coleman1985q, sakai2008stability}. This process of the formation of Q-balls in the thermal history of the Universe is termed solitosynthesis \cite{postma2002solitosynthesis}, just like the case of nucleosynthesis. Solitosynthesis of large Q-balls in the RD epoch in thermal equilibrium is only possible if freeze-out does not hinder the following chain of reactions leading from small to large solitons. Unlike some other non-topological solitons, the classically stable Q-balls must exist for very small charges $Q \geq 1$ (For a detailed discussion, see Ref.~\cite{kusenko1997phase}):
\begin{equation}
\begin{split}
 \Phi + \Phi & \rightarrow (Q = 1)\\ 
  & \vdots\\
 (Q) + \Phi & \rightarrow (Q + 1)\\
\end{split}
\end{equation}

In principle, the Q-balls can form with varied charge ($Q$) values. However, the Q-balls, which slowly dominate over the radiation, depend on their charge $Q$. Therefore, knowing the RD-MD equality time, we can, in principle, fix the value of the mass of the Q-balls, which also depends on the charge $Q$. We show this explicitly in the next section. With this input, we can constrain the mass of Q-balls, which can be treated as a DM candidate.

The above physical inputs suggest that the best scenario for the Q-ball formation in the RD epoch before the matter-radiation equality, where the energy scale is around $1~{\rm eV}$. Comparing the above energy scale with $\bar{T}_c$, we get $m_{\sigma} \sim 10$~KeV. This corresponds to the fact that the above transition occurs during the RD epoch, particularly during the recombination time scale, since, at this time, charged particles start forming heavier elements. In other words, once such phase transition occurs above the critical temperature $\bar{T}_{c} = T_{c}/m_{\sigma} > 1$, the composite Q-balls start forming through the accumulation of constituent charged particles.

Although the total global $U(1)$ charge of  Q-ball is a free parameter, we can obtain a bound on this value using the constraint from the invisible decay mode of ortho-positronium~\cite{PhysRevD.88.117701, PhysRevD.75.032004, PhysRevD.75.075014}. Since the Q-ball mass is much smaller than the electron's mass, the invisible decay mode of ortho-positronium leads to $Q < 3.4 \times 10^{-5}$~\cite{PhysRevD.88.117701, PhysRevD.75.032004, PhysRevD.75.075014}. For thin-wall Q-ball in the present epoch, Eqs. (\ref{eq. 35}) and \eqref{eq. 38} lead to:
\begin{equation}
\label{eq:constraint02}
\begin{split}
\mathcal{E}_{\lambda_{0}} & = Q\sqrt{m_{\sigma}^{2} - \frac{\zeta}{2}\sigma_{0}^{2}
 + \frac{\xi}{3}\sigma_{0}^{4}}\\
%
\implies Q^{ - 1} & = m_{\sigma}\sqrt{1 - \frac{3\zeta^{2}}{16\bar{\xi}}} > 
3 \times 10^{4}\text{eV}.   
\end{split}
\end{equation}
From the above expression and Eq.~\eqref{eq:constraint01}, we see that $m_{\sigma} > 30$~KeV. Interestingly, the ortho-positronium constraint leads approximately to the same value of $m_\sigma$ obtained from a completely different physical condition above. Based on the experimental data, this consistency check bolsters the millicharged Q-balls with mass less than 1~eV as a possible dark matter candidate in the cosmological scales.

Interestingly, the density perturbations that reenter during the recombination epoch increase the production of Q-balls. The density perturbations within the Hubble radius during the RD epoch can be expressed as temperature fluctuations, i.e., 
\[
T(t,\vec{x}) = T_{0}(t) + \delta T(t,\vec{x})
\] 
where $T_{0}$ is the ambient temperature and $\delta T$ represents the temperature fluctuations in the Universe. Thus, when the Universe expands, and the homogeneous temperature of the Universe decreases as the scale factor, the inhomogeneities in local regions can be above the critical temperature $T_c$ for Q-ball formation. This can be achieved if $\delta T(t,\vec{x})$ at a given region is such that $T(t,\vec{x})$ is above the critical temperature. Thus, the scalar perturbations generated during inflation that left the horizon around the last 5 to 10 e-foldings of inflation and reentered in the RD epoch provide the physical mechanism for the Q-ball formation. As a result, more and more Q-balls start forming. Since the Q-ball is a stable configuration, the charged particles corresponding to the complex scalar field $\Phi$ will prefer to interact with Q-balls. Hence, effectively, Q-balls are decoupled from the photons. Once the Q-balls and other normal matter start forming, the Universe transitions toward the matter-dominated era. Our proposal differs from the earlier approach of Q-ball formation in the early universe~\cite{1989-Frieman.etal-PRD}. Specifically, our proposal for Q-ball formation depends on the temperature in a given overdense region. Hence, for the mass $m_{\sigma}$, we can have multiple Q-balls formed in various regions, and as discussed in Appendix A, they are stable. Appendix B shows that, in the thin-wall approximation, the Q-ball approximation has more energy.

Having shown that the Q-balls can form in the recombination epoch, we need to confirm whether these Q-balls are indeed stable. In Appendix~\eqref{app:Derricktheorem}, we use Derrick's theorem to show that these Q-balls are indeed stable.

\section{Q-balls as ``Cold" Dark matter in Cosmological scales}\label{section 6}
In the previous section and Appendix \eqref{app:Derricktheorem}, we have established that the Q-balls can form in the RD epoch, and these Q-balls are stable. As shown in Appendix \eqref{app:thinwall}, in the thin-wall approximation, the core value of the field ($\sigma_0)$ increases. This implies that the energy of the Q-ball increases during the RD epoch. This raises the question: Can the Q-balls be considered dark matter? If yes, can they have a sufficient energy density that triggers the transition from RD epoch to Q-matter dominated epoch in a finite time? It then remains to be determined whether these constraints are consistent with the observations and theoretical constraints we obtained in the previous section. 

In Appendix \eqref{app:thinwall}, under thin-wall approximation, we obtained the decay of energy density of Q-ball in the RD epoch to be
\begin{equation}
\rho_{Q} \sim \frac{1}{a^{3}(\eta)}
\end{equation}
This is identical to the energy density of pressure-less matter decay. Hence, Q-balls can act as dark matter and in cosmological scales can be treated as CDM which results in a transition from the RD epoch to the MD epoch. This result is crucial for establishing that the Q-balls can act as dark matter. In Sec.~\eqref{sec:RDtoQMD}, we show this explicitly.

In the rest of this section, we address the above questions by considering the possibility of Q-balls as a dark matter candidate and its implications in standard cosmology.

\subsection{Rate of formation of Q-balls and its number density in thermal equilibrium}
After the formation of Q-balls, the photon decouples once the Universe becomes matter-dominated. In this epoch, we can treat Q-balls as composite particles or objects in thermodynamic equilibrium with the free $\Phi$ particles. More specifically, the following reaction is in equilibrium:
\begin{equation}\label{eq. 51}
(Q) + \Phi \leftrightarrow (Q + 1) \, ,
\end{equation}
and the chemical potentials of Q-balls ($\mu_{Q}$) and free $\Phi$ ($\mu_{\Phi}$) particles are related as $\mu_{Q} = Q\mu_{\Phi}$.

These particles can be treated as non-relativistic since $T < m_{\sigma}$, hence, the number densities of Q-balls and free $\Phi$ particles can be approximated by the Boltzmann distribution:
\begin{equation}\label{eq. 52}
f_{BE}(p) = \frac{1}{e^{\beta(E_{\sigma}(P) - \mu_{\Phi})} - 1}\approx e^{(\mu_{\Phi} - 
m_{\sigma})/T}e^{ - \frac{p^{2}}{2m_{\sigma}T}}.
\end{equation}
Thus, the number density for the Q-balls and free $\Phi$ particles, respectively, are
\begin{eqnarray}
\label{eq. 54}
n_{Q} &=& g_{Q}\left(\frac{E_{Q}T}{2\pi}\right)^{3/2}e^{(\mu_{Q} - E_{Q})/T}, \\
\label{eq. 53}
n_{\Phi} &=& g_{\Phi}\left(\frac{m_{\sigma}T}{2\pi}\right)^{3/2}e^{(\mu_{\Phi} - m_{\sigma}) /T}. 
\end{eqnarray} 
Here $g_{\Phi} = 2$ and $g_{Q}$ is the internal partition function of Q-ball~\cite{postma2002solitosynthesis}. Using the fact that free particles $\Phi$ and Q-balls are thermal equilibrium, the number density of Q-balls can be rewritten as:
\begin{equation}\label{eq. 55}
n_{Q} = g_{Q}\left(\frac{E_{Q}T}{2\pi}\right)^{3/2}e^{(Q\mu_{\Phi} - E_{Q})/T}.
\end{equation}
As a result, we obtain the following relation between the two number densities
\begin{equation}\label{eq. 56}
\begin{split}
\frac{n_{Q}}{n_{\Phi}^Q} & = \frac{g_{Q}}{2^{Q}}\left(\frac{E_{Q}}{m_{\sigma}}\right)^{3/2}
\left(\frac{2\pi}{m_{\sigma}T}\right)^{3(Q - 1)/2}e^{B_{Q}/T}\\
 & = Q^{3/2}\frac{g_{Q}}{2^{Q}}\left(\frac{\lambda_{0}}{m_{\sigma}}\right)^{3/2}\left(\frac{2\pi}{m_{\sigma}T}\right)^{3(Q - 1)/2}e^{B_{Q}/T},
\end{split}
\end{equation}
where $B_{Q} = Qm_{\sigma} - E_{Q} = Q(m_{\sigma} - \lambda_{0}) > 0$ is the binding energy of the Q-ball. In deriving the above condition, we have used the thin-wall approximation relation \eqref{eq:thinwall-EQ}. 
While the results in Appendix B are obtained for the RD epoch, the analysis is usually valid for the MD epoch. The quantity on the left-hand side is bounded from above by
\begin{equation}\label{eq. 57}
\frac{n_{Q}}{n_{\Phi}^{Q}} < Q^{3/2}\frac{g_{Q}}{2^{Q}}\left(\frac{2\pi}{m_{\sigma}T}\right)
^{3(Q - 1)/2}e^{B_{Q}/T}.
\end{equation}
The above expression determines the size of the Q-ball in order for the Q-balls to be in thermal equilibrium with free $\Phi$ particles. 

Having obtained this, we can now evaluate the rate of Q-ball formation for the above reaction \eqref{eq. 51}. Specifically, we are interested in a $Q$ unit of charge combined with a $\Phi$ particle, forming a Q-ball of charge $Q + 1$ unit and its reverse.
The following Boltzmann's formula can express the reaction rate for this process in the cosmological setting:
\begin{equation}\label{eq. 66}
\frac{1}{a^{3}}\frac{d(n_{Q}a^{3})}{dt} = - \alpha n_{Q}n_{\Phi} + \beta n_{Q+1},
\end{equation} 
where $\alpha = \langle\sigma v\rangle$ is the thermally-averaged collisional cross-section 
which comes from the collision term in the Boltzmann equation. On the other hand, $\beta$ can be obtained from the thermal equilibrium condition, i.e.,
%
\begin{equation}\label{eq. 67}
\beta = \alpha\left(\frac{n_{Q}n_{\Phi}}{n_{Q + 1}}\right)_{eq}.
\end{equation} 
Substituting these in Eq.~(\ref{eq. 66}), we have:
\begin{equation}\label{eq. 70}
\frac{1}{n_{Q}a^{3}}\frac{d(n_{Q}a^{3})}{d\log a} = - \frac{\Gamma}{H}\Big[1 - \frac{n_{Q + 1}}
{n_{Q}n_{\Phi}}\left(\frac{n_{Q}n_{\Phi}}{n_{Q + 1}}\right)_{eq}\Big].
\end{equation}
where 
\begin{equation}\label{eq. 69}
\Gamma = n_{\Phi}\langle\sigma v\rangle \, .
\end{equation}
From the above expression, we see that $(\Gamma/H) \sim (1/T)$ in the RD epoch, while in the MD epoch, we have $(\Gamma/H) \sim 1/T^{1/2}$. This implies that the prefactor in the RHS of Eq.~\eqref{eq. 70} decays faster \textit{w.r.t} temperature in the RD epoch compared to the MD epoch. 
This is because the temperature in the Universe during the MD epoch is much smaller compared to the RD epoch. One may re-express $\Gamma/H$ in terms of $T_{0}/T(z)$ where $T_{0}$ is the current temperature of the Universe and $T(z)$ is the temperature at a given epoch.

The factor with `eq' subscript inside the parenthesis on the right-hand side can be obtained from the equilibrium number density expressions mentioned previously, and it
is of the following form
\begin{equation}\label{eq. 71}
\begin{split}
\left(\frac{n_{Q}n_{\Phi}}{n_{Q + 1}}\right)_{eq} & = \frac{g_{Q}g_{\Phi}}{g_{Q + 1}}\left(
\frac{m_{\sigma}E_{Q}T}{2\pi E_{Q + 1}}\right)^{3/2}e^{[E_{Q + 1} - m_{\sigma} - E_{Q}]/T}\\
 & = \frac{g_{Q}g_{\Phi}}{g_{Q + 1}}\left(\frac{m_{\sigma}QT}{2\pi(Q + 1)}\right)^{3/2}
 e^{- \Delta/T}\\
 & < \frac{g_{Q}g_{\Phi}}{g_{Q + 1}}\left(\frac{m_{\sigma}T}{2\pi}\right)^{3/2}e^{- \Delta/T}.
\end{split}
\end{equation}
where we have used the thin-wall approximation discussed in Appendix \eqref{app:thinwall} to rewrite $E_Q$ and $\Delta = m_{\sigma} - \lambda_0$. Note that in the MD era, temperature $T \ll \Delta$ and the above upper bound are exponentially suppressed. However, in the RD epoch, when $T \gg \Delta$, the upper bound grows as $T^{3/2}$. This shows that as the temperature drops, the Q-ball formation decreases. It is also important to note that as the temperature goes down as the Universe expands, the interaction rate also goes down. It eventually becomes much smaller compared to Hubble rate $H$, at which point $n_{Q}\sim 1/a^{3}$.

Having obtained the rate of Q-ball formation for the above process \eqref{eq. 51}, we are now interested in evaluating the number density of Q-balls at the MD epoch. This is crucial to determine whether the Q-balls can act as dark matter in the early universe and seed structure formation. In order to do this, it is necessary to compare it with the number density of photons, as they are present now and were present in the early Universe. As discussed in the previous section, the charge asymmetry parameter $(\eta_{Q})$, a conserved quantity, provides a crucial link to compare with photons. Substituting Eq.~\eqref{eq. 62} in Eq.~\eqref{eq. 57}, we have:
\begin{equation} \label{eq. 64}
\begin{split}
\frac{n_{Q}}{n_{\gamma}^{Q}} & < Q^{3/2}(\eta_{Q})^{Q}\frac{g_{Q}}{2^{Q}}\left(\frac{2\pi}{m_{\sigma}
T}\right)^{3(Q - 1)/2}e^{B_{Q}/T}\\
\frac{n_{Q}}{n_{\gamma}^{(0)Q}} & < Q^{3/2}(\eta_{Q})^{Q}\frac{g_{Q}}{2^{Q}}\left(\frac{2\pi}{m_{\sigma}
T_{0}}\right)^{3(Q - 1)/2}e^{aB_{Q}/T_{0}}\frac{1}{a^{3(Q + 1)/2}}\\
\frac{n_{Q}}{n_{\gamma}^{(0)Q}} &< g_{Q} \left(Q\frac{m_{\sigma}T_{0}}{2\pi a}\right)^{3/2}
\Big[\eta_{Q}\left(\frac{2\pi}{m_{\sigma}T_{0}a}\right)^{3/2}\frac{e^{a\Delta/T_{0}}}{2}
\Big]^{Q}.
\end{split}
\end{equation}
where $\Delta = m_{\sigma} - \lambda_{0} > 0$, $T_{0}$ refers to the present temperature of the Universe and $n_{\gamma}^{(0)}$ corresponds to the present density of photons in the Universe. From the above expression, we can infer that, for a given non-zero value of $\eta_{Q}$, $n_{Q}$ gets larger and larger in the past for large $Q$ values. Alternatively, as we go back in time, the number density of the Q-ball grows.

Our analysis suggests that millicharged composite Q-balls can be considered a dark matter candidate for the following reasons. First, they are weakly interacting and stable non-topological soliton configurations, which means they do not interact with normal matter strongly. Second, their formation in the RD epoch depends on the phase transition that occurs in distinct places in the Universe due to temperature fluctuations. Hence, they have a non-trivial distribution in the Universe. We may also note that temperature fluctuations in the RD epoch are related to the matter density fluctuations; hence, there is a strong correlation between dark matter distribution and the late-time structure formation of the Universe. 
Lastly, following the expressions of the number density of Q-balls produced in MD epoch Eq. (\ref{eq. 54}) and number density of $\Phi$ particles Eq. (\ref{eq. 53}), we can compute the following ratio
\begin{equation}
\frac{n_{Q}}{n_{\Phi}} = \frac{g_{Q}}{g_{\Phi}}\left(\frac{E_{Q}}{m_{\sigma}}\right)^{3/2}
\exp\left(\frac{m_{\sigma} - E_{Q}}{T} + \frac{\mu_{Q} - \mu_{\Phi}}{T}\right).
\end{equation}
Since $m_{\sigma} \gg E_{Q}$ and $\mu_{Q} = Q\mu_{\Phi} \ll \mu_{\Phi}$, the above expression reduces to:
\begin{equation}
\begin{split}
\frac{n_{Q}}{n_{\Phi}} &\sim \frac{g_{Q}}{g_{\Phi}}\left(\frac{E_{Q}}{m_{\sigma}}\right)^{3/2}
\exp\left( - \frac{\mu_{\Phi} - m_{\sigma}}{T}\right) \\ 
& \sim 10^{-6}  \times\exp\left( - \frac{\mu_{\Phi} - 
m_{\sigma}}{T}\right) \, .
\end{split}
\end{equation}
Since $\mu_{\Phi} > m_{\sigma}$, the exponent is decaying function and in the MD epoch is close to unity. This shows that the number density of Q-balls in MD epoch is very small compared to the number density of $\Phi$-particles. This shows that if Q-balls can explain DM at cosmological scale then the number density of $\Phi$-particles should be atleast $10^{6}$ times larger compared to the number density of Q-balls. Since $m_{\sigma} > m_{Q-ball}$, the energy required to produce Q-ball is much less than the energy available in the $\Phi$ field. Hence, the mechanism to produce Q-balls in the early Universe is viable.

Therefore, for the above-mentioned strong reasons, we propose that Q-balls can be considered a dark matter candidate. We now compare the theoretical and observational constraints for the Q-ball configurations to drive from the RD epoch to the MD epoch.

\subsection{Radiation-dominated era to Q-matter-dominated era transition}
\label{sec:RDtoQMD}
In the previous subsection, we have shown that the number density of Q-balls grows as we go back in time. Given this, we now obtain the condition for the transition from RD epoch to Q-matter dominated epoch. 

Photons and neutrinos were the dominant form of energy in the early Universe and are given by
\begin{equation}\label{eq. 72}
\rho_{\rm rad} = \left[1 + \frac{7 N_{\nu}g_{\nu}}{8}\left[\frac{4}{11}\right]^{4/3}\right]
\rho_{\gamma},~~~
\rho_{\gamma} = 
\frac{\pi^{2}T_{0}^{4}}{15a^{4}} \, ,
\end{equation}
where $N_{\nu}$ and $g_{\nu}$ are the number of degrees of neutrinos and the degeneracy of
states for neutrinos, respectively. 
The energy density of Q-balls of charge $Q$ unit is:
\begin{equation}\label{eq. 74}
\begin{split}
\rho_{Q} & = E_{Q}n_{Q} = g_{Q}\left(\frac{E_{Q}T}{2\pi}\right)^{5/2}e^{(\mu_{Q} - E_{Q})/T}
\frac{2\pi}{T}\\
 & = g_{Q}\left(\frac{M_{Q}T}{2\pi}\right)^{5/2}\frac{2\pi}{T}e^{Q(\mu_{\Phi} - \lambda_0)/T} \\
& = g_{Q}\left(\frac{M_{Q}T_{0}}{2\pi}\right)^{5/2}\frac{2\pi}{T_{0}}\frac{1}{a^{3/2}}
 e^{aQ\Delta/T_{0}} 
\end{split}
\end{equation}
In deriving the last expression, we have used the relation between $\mu_Q$ and $\mu_{\Phi}$ in the previous subsection and $\mu_{\Phi}\sim m_{\sigma}$. 
To obtain the epoch in which the radiation and Q-matter equality happens, we equate the above expression with Eq. \eqref{eq. 72}, i.e., $\rho_Q = \rho_{\rm rad}$. This leads to:
\begin{equation}\label{eq. 76}
\begin{split}
g_{Q}\left(\frac{E_{Q}T_{0}}{2\pi}\right)^{5/2} & \frac{2\pi}{T_{0}}\frac{1}{a^{3/2}} \simeq
\left(1 + \frac{7}{8}N_{\nu}g_{\nu}\left(\frac{4}{11}\right)^{4/3}\right)\frac{\pi^{2}
T_{0}^{4}}{15a^{4}}\\
\implies a^{5/2} & = \frac{2\pi}{E_{Q}}\mathcal{K}_{Q}^{2/5}T_{0} = \frac{2\pi\mathcal{K}_{Q}^{2/5}}{Q}
 \frac{T_{0}}{\lambda_0},
\end{split}
\end{equation}
where
\begin{equation}\label{eq. 77}
\mathcal{K}_{Q} = \left(1 + \frac{7}{8}N_{\nu}g_{\nu}\left(\frac{4}{11}\right)^{4/3}\right)
\frac{\pi}{30g_{Q}}.
\end{equation}
Hence, by viewing Q-matter as a potential candidate for dark matter and requiring the equality of radiation and matter at a redshift of $z = 3600$, it is possible to impose constraints on either the values of charge $Q$ and $\lambda_{0}$ or the mass $E_{Q}$. Second, using that fact that $T_0 = 2.7$ Kelvin and $g_{\nu} N_{\nu}= 3$, we get:
\begin{eqnarray}
E_{Q} & =&  2\pi T_{0}(1 + z_{eq})\Bigg[\left(1 + \frac{21}{8}\left(\frac{4}{11}\right)^{4/3}
\right)\frac{\pi}{30g_{Q}}\Bigg]^{2/5} \nonumber \\
\label{eq:Qball-Mass}
& =& 2\pi T_{0}(1 + z_{eq})\left(\frac{0.176}{g_{Q}}\right)^{2/5} = \frac{2.6}{g_{Q}^{2/5}}
 \text{eV}.
 \end{eqnarray}
The expression above serves as an upper limit for the mass of Q-balls, as it takes into account the component $e^{aQ\Delta/T_{0}} > 1$. 

Within the thin-wall approximation, the mass (or co-moving energy of the associated static configuration) of Q-balls with charge Q in our model is determined by (see Eq. (\ref{eq:thinwall-EQfin})):
\begin{equation}
M_{Q} \equiv \frac{E_{Q}}{a} = Q\sqrt{m_{\sigma}^{2} - \frac{3\zeta^{2}}{16\xi}}, 
\end{equation}
Their co-moving radius is given by (see Eq. (\ref{Co-moving radius of Q-balls})):
\begin{equation}
aR_{Q} \sim \left(\frac{Q\xi}{\zeta\sqrt{m_{\sigma}^{2} - \frac{3\zeta^{2}}{16\xi}}}\right)
^{1/3} \, ,
\end{equation}
Consequently, in natural units, the Compton wavelength of these Q-balls is:
\begin{equation}
\frac{1}{M_{Q}} = \frac{1}{Q\sqrt{m_{\sigma}^{2} - \frac{3\zeta^{2}}{16\xi}}}.
\end{equation}
This leads to the following ratio:
\begin{equation}
\frac{\text{Compton wavelength}}{\text{Co-moving radius}} = \frac{1/M_{Q}}{aR_{Q}} \sim
\frac{1}{Q^{4/3}}\frac{1}{\left(\frac{m_{\sigma}^{2}\xi}{\zeta} - \frac{3\zeta}{16}\right)
^{1/3}}.
\end{equation}
It is important to emphasize that these results hold true at any epoch in the cosmic history as we have evaluated comoving quantities. It is important here to note that the equation (\ref{eq. 25 revised draft}) establishes the necessary condition for Q-ball stability, ensuring they do not decay into $\Phi$ quanta. Importantly, in the present Universe, $\zeta$ is extremely small and approaches zero. This significantly weakens the attractive part of the field potential, inhibiting Q-ball formation. As a result, both the Compton wavelength and thermal de Broglie wavelength\footnote{Thermal de Broglie wavelength $\lambda_{th}$ denotes the inter-particle separation between the scalar field quantas in thermal equilibrium.} become smaller than the coherence length or the size of the Q-balls in the current epoch.

%

\section{``superfluid" Q-Balls in galactic scales}
\label{section 7}

On the galactic or sub-galactic structure scale, the $\Lambda$CDM model --- cold dark matter (CDM) with the cosmological constant --- seems to encounter difficulties despite its widespread use and remarkable success in explaining the large-scale structure of the Universe \cite{khoury2022dark, berezhiani2015theory}. Contrary to observations, numerical simulations with the $\Lambda$CDM model typically predict a cusped central density and too many sub-halos \cite{khoury2022dark, berezhiani2015theory}. Since typical CDM particles, such as WIMPs, are heavy and slow, they tend to cluster together to form smaller structures, such as a dark matter star or planet. In contrast, observations suggest that the smallest DM-dominated structure is a dwarf galaxy \cite{Lee:2008ab}. This implies that DM has a natural minimum length scale. In addition, observed halo spins appear to be larger than those predicted by the $\Lambda$CDM model. Although the significance of this disparity remains debatable, it is imperative to look for alternatives or a unified framework. Consequently, it is desirable to contemplate an alternative DM candidate that can function as a CDM on scales larger than a galaxy while suppressing sub-galactic structures. 

This section shows that Q-balls can be a good candidate. In other words, on the largest length scale, Q-balls can behave like CDM in cosmological scales; however, at late times, due to the self-interacting potential, they can form BEC and can have interesting implications on small scales. 

\subsection{Bose-Einstein condensation by Q-balls}

In Appendix \ref{app:BEC}, we obtain two conditions for forming BEC. The two conditions in the context of DM can be translated as follows: DM particles must be Bosons, and their thermal de Broglie wavelength must overlap with the inter-particle separation within the galaxies. 
This leads us to the following two questions which we try to answer in the rest of this work: What must be the mass of the DM particles in order to form the BEC? How does the Q-balls fit into this?

To answer the first question, we ascribe an effective DM temperature for the DM particles moving with a velocity $v$ to be:
\[
\frac{3}{2}k_{B}T = \frac{1}{2}mv^{2} \, .
\]
Thus, the thermal de Broglie wavelength is 
\begin{equation}\label{eq. 2.49}
\lambda_{\rm th} = \frac{\sqrt{6\pi}}{mv}.
\end{equation}
and the condition for BEC translates to 
\begin{equation}\label{eq. 2.50}
\frac{1}{mv} \geq \frac{\zeta(3/2)}{\sqrt{6\pi}}n^{-1/3} \simeq 0.6 n^{-1/3} = 0.6\left(
\frac{m}{\rho}\right)^{1/3}.
\end{equation}
As mentioned earlier, in the current epoch, $\zeta$ is extremely small and approaches zero. This significantly weakens the attractive part of the field potential, inhibiting Q-ball formation. As a result, both the Compton wavelength and thermal de Broglie wavelength become smaller than the coherence length or the size of the Q-balls in the current epoch. Since we are considering the condensation of scalar field quanta in the present epoch into the ground state, which is effectively the coherent state formed by the Q-ball, it is sufficient to compare de Broglie wavelength.

Given the energy density $\rho = mn$ and velocity dispersion in the neighborhood of the Milky Way galaxy, the above relation puts an upper bound on the mass of DM particles that can form BEC:
\begin{equation}\label{eq. 2.51}
m \leq \left(\frac{4.63\rho}{v^{3}}\right)^{1/4}.
\end{equation}
The above expression answers whether BEC formation (by Q-balls) can occur at the galaxy scales. To go about this, we consider the Milky way galaxy to be typical in the current Universe. Taking the local average density of the Milky Way galaxy to be $\rho \simeq 10^{-25}g/cm^{3} = 4\times 10^{-7}eV^{4}$ \cite{sofue2012grand} and the velocity of the dark matter particles to be $v = 10^5 m/s$~\cite{Freese:2012xd}, we obtain the following bound on the particle's mass $m \leq 15 eV$ in order to form BEC. In other words, any particle whose mass is less than $15~{\rm eV}$ can form a BEC in the galaxy scale. From Eq.~\eqref{eq:Qball-Mass}, we infer that the mass of Q-balls, formed in the RD epoch and drives the Universe to the MD epoch, is of the order eV. 

\subsection{Dark matter superfluid}\label{section 8}
As discussed in the Introduction,  Bose-Einstein condensation and superfluidity are intimately related. In order to acquire a superfluid phase, Bose-Einstein condensation must first occur; however, the converse is not true, as the superfluidity property disappears in the absence of interactions. The superfluidity of DM dramatically affects the macroscopic behavior of DM in galaxies. Rather than acting as individual collisionless particles, DM can be more accurately described as a collective excitation, which, at low energies, are only Phonons. It is well known that superfluid particles are generally described by the scalar field, governed by the effective field theory, as described in the Appendix. \ref{App:superfluid}.
These phonons are essential in mediating a long-range force between ordinary (baryonic) matter particles \cite{berezhiani2015theory}. As a result, any test particle orbiting the galaxy is now subject to two kinds of forces: the Newtonian gravitational force and the phonon-mediated force~\cite{khoury2022dark, berezhiani2015theory}.

In the previous subsection, we showed that the Q-balls can form BEC. This leads us to the following question: Can Q-balls form a superfluid phase? Since the average temperature of the present Universe is $2.7K$, we neglect the finite-temperature corrections in the analysis. In the following subsection, we explicitly show that the Q-balls condensate formed in the early Universe can form a superfluid phase at late times due to contact repulsive interaction. Specifically, assuming that the Milky Way galaxy is typical in the current Universe, we obtained expression \eqref{eq. 2.73} that determines when the objects passing through the DM superfluid medium feel friction. Interestingly, for a Q-ball dark matter of mass $10$~eV, we obtain the critical velocity below which the superfluidity phenomenon can be observed. 

\subsection{Dark Matter superfluidity}\label{App:superfluid}

The simplest model of superfluidity is described by the following Lagrangian density
\begin{equation}\label{eq. 2.54}
\mathcal{L} = - \partial_{\mu}\mathcal{Q}^{*}\partial^{\mu}\mathcal{Q} - m^{2}|\mathcal{Q}|^{2} - \frac{g}{2}|\mathcal{Q}|^{4},
\end{equation}
where $g$ is the self-interaction coupling which is dimensionless. The above Lagrangian density is invariant under the global $U(1)$ symmetry $\mathcal{Q}\rightarrow\mathcal{Q}e^{i\alpha}$ where $\alpha$ is a constant. Therefore, according to Noether's theorem, the associated conserved 4-current is expressed by
\begin{equation}\label{eq. 2.55}
j^{\mu} = 2\text{Im}(\mathcal{Q}^{*}\partial^{\mu}\mathcal{Q}),
\end{equation}
whereas the conserved charge or particle number density is expressed as
\begin{equation}\label{eq. 2.56}
n = - 2\text{Im}(\mathcal{Q}^{*}\dot{\mathcal{Q}}).
\end{equation} 
In order to describe the condensate in the superfluid phase, we consider the mean-field theory in which the condensate wavefunction $\mathcal{Q}_{0}$ satisfies the following classical equation of motion
\begin{equation}\label{eq. 2.57}
\partial^{2}\mathcal{Q}_{0} = m^{2}\mathcal{Q}_{0} + g|\mathcal{Q}_{0}|^{2}\mathcal{Q}_{0}.
\end{equation}
Assuming the homogeneous condensate in its rest frame, we consider the following ansatz for the condensate wavefunction
\begin{equation}\label{eq. 2.58}
\mathcal{Q}_{0}(t) = v e^{i\mu_{R}t}
\end{equation}
where $\mu_{R}$ is the relativistic chemical potential. Substituting the above ansatz into the equation (\ref{eq. 2.57}), we obtain the following relation
\begin{equation}\label{eq. 2.59}
\mu_{R}^{2} = m^{2} + gv^{2},
\end{equation}
which in the non-relativistic limit becomes the following
\begin{equation}\label{eq. 2.60}
\mu_{R} \simeq m + \frac{gv^{2}}{2m}.
\end{equation}
We use the above-mentioned non-relativistic result, as the DM particles are believed to be non-relativistic in nature. Therefore, we introduce the non-relativistic chemical potential by
\begin{equation}\label{eq. 2.61}
\mu \equiv \mu_{R} - m = \frac{gv^{2}}{2m}.
\end{equation}
Substituting the ansatz (\ref{eq. 2.58}) in the equation (\ref{eq. 2.56}), we obtain the following relation
\begin{equation}\label{eq. 2.63}
n = 2\mu_{R}v^{2} = 2mv^{2},
\end{equation}
which can also be expressed as
\begin{equation}\label{eq. 2.64}
n \simeq \frac{4m^{2}\mu}{g}.
\end{equation}
On the other hand, according to Goldstone's theorem, we expect the existence of a massless/gapless excitation due to the spontaneous symmetry breaking of global $U(1)$ symmetry, and these gapless excitations are the phonons. In order to describe the system, let us take into account the fluctuations into account which make the $\mathcal{Q}$ field as follows
\begin{equation}\label{eq. 2.65}
\mathcal{Q}(\vec{x}, t) = (v + h(\vec{x}, t))e^{i(\mu_{R}t + \pi(\vec{x}, t))}.
\end{equation} 
Substituting the above into the Lagrangian density (\ref{eq. 2.54}), we obtain the following Lagrangian density
\begin{equation}\label{eq. 2.66}
\mathcal{L} = - \partial_{\mu}h\partial^{\mu}h + (v + h)^{2}[gv^{2} + 2\mu_{R}\dot{\pi} + 
\dot{\pi}^{2} - (\vec{\nabla}\pi)^{2}] - \frac{g}{2}(v + h)^{4}.
\end{equation}
Expanding the above Lagrangian density to quadratic order in fields, we find that the mass of $h$ field is given by $m_{h}^{2} = 2gv^{2}$. Since the $h$ field is massive in nature, in the low energy/momentum limit, it is justified to integrate out the field $h$ order by
order in its derivatives. To the zeroth order within the saddle point approximation, we have the following equation of motion
\begin{equation}\label{eq. 2.67}
g(v + h)^{2} = gv^{2} + 2\mu_{R}\dot{\pi} + \dot{\pi}^{2} - (\vec{\nabla}\pi)^{2} \equiv
X_{R}.
\end{equation}
Substituting the above relation into the action (\ref{eq. 2.66}), we obtain the following effective action for the phonons
\begin{equation}\label{eq. 2.68}
\mathcal{L}_{\pi} = \frac{1}{2g}X_{R}^{2},
\end{equation}
where $X_{R}$ in non-relativistic limit ($\mu_{R} = m, \ \dot{\pi} \ll m$) reduces to the following
\begin{equation}\label{eq. 2.69}
X_{R} \simeq 2m\left(\mu + \dot{\pi} - \frac{(\vec{\nabla}\pi)^{2}}{2m}\right) \equiv 2m X.
\end{equation}
Therefore, at zero temperature in the non-relativistic limit, effective action for the gapless mode $\pi$ is of the following form
\begin{equation}\label{eq. 2.70}
\mathcal{L}_{\pi} = \frac{2m^{2}}{g}X^{2},
\end{equation}
which is invariant under the shift symmetry $\pi \rightarrow \pi + \alpha$ which is the original $U(1)$ symmetry. To the quadratic order, $\mathcal{L}_{\pi}$ reduces to the following form
\begin{equation}\label{eq. 2.71}
\mathcal{L}_{\pi} \simeq \frac{2m^{2}}{g}\mu^{2} + \frac{4m^{2}\mu}{g}\dot{\pi} + 
\frac{2m^{2}}{g}\left(\dot{\pi}^{2} - \frac{\mu}{m}(\vec{\nabla}\pi)^{2}\right),
\end{equation}
where the first two terms can be dropped as they are constant and first-order time-derivative terms. The dispersion of these gapless excitations is of the following form
\begin{equation}\label{eq. 2.72}
\omega = c_{s}k, \ c_{s} = \sqrt{\frac{\mu}{m}} = \sqrt{\frac{gn}{4m^{3}}} = 
\frac{\sqrt{g\rho}}{2m^{2}}.
\end{equation}
Considering the local average energy density of the Milky Way galaxy $\rho$ to be $4\times 10^{-7}eV^{4}$ and $m = 10 eV$, we obtain the following expression of speed of sound $c_{s}$
\begin{equation}\label{eq. 2.73}
c_{s} \simeq 9.5\sqrt{g}~Km/s.
\end{equation}
Depending on the value of $g$, the speed of sound determines 
when the objects passing through the  DM superfluid medium feel friction. Specifically, objects moving with velocity $v < c_s$ will not feel any friction and objects moving with velocity $v > c_s$ feel the friction since they can create phonons.

We may note that in the absence of phonons, the bosonic condensate at $T = 0$ leads to a
non-zero pressure which is expressed as
\begin{equation}\label{eq. 2.74}
P_{cond} = \frac{2m^{2}}{g}\mu^{2},
\end{equation}
whereas its number density is given by
\begin{equation}\label{eq. 2.75}
n = \frac{\partial P}{\partial\mu} = \frac{4m^{2}}{g}\mu.
\end{equation}
Therefore, the equation of state associated with the condensate of the superfluid state is 
given by
\begin{equation}\label{eq. 2.76}
P_{cond} = \frac{g}{8m^{2}}n^{2} = \frac{g}{8m^{4}}\rho_{cond}^{2}.
\end{equation}
As a result, the continuity equation of DM superfluid condensate becomes
\begin{equation}\label{eq. 2.77}
\dot{\rho}_{cond} + 3H\left(\rho_{cond} + \frac{g}{8m^{4}}\rho_{cond}^{2}\right) = 0,
\end{equation}
whose solution is given by
\begin{equation}\label{eq. 2.78}
\left(1 + \frac{8m^{4}}{g\rho_{0}}\right)a^{3} = 1 + \frac{8m^{4}}{g\rho_{cond}},
\end{equation}
which is $g\rightarrow 0$ limit leads to a pressure-less fluid solution. However, in the presence of a weak interaction that produces the superfluid phase, we obtain the pressure of the DM superfluid condensate to be non-zero. Moreover, the energy density of condensate
evolves as
\begin{equation}\label{eq. 2.79}
\rho_{cond} = \frac{1}{\left(\frac{g}{8m^{4}} + \frac{1}{\rho_{0}}\right)a^{3} - 
\frac{g}{8m^{4}}},
\end{equation}
where $\rho_{0} = \rho_{cond}(t = 0)$.
Given this setup, we are now in a position to show that the Q-balls formed in the early Universe can lead to superfluid dark matter in the late universe. 

\subsection{MOND from dark matter superfluid}\label{MOND from DM superfluid}

Like in section \ref{section 1}, now we consider our model of Q-ball matter in the present epoch, which is described by the following Lagrangian density:
\begin{equation}
\mathcal{L}_{Q} = - \partial_{\mu}\Phi\partial^{\mu}\Phi^{*} - m^{2}|\Phi|^{2} + \frac{\zeta}{2}
|\Phi|^{4} - \frac{\xi}{3}|\Phi|^{6}.
\end{equation}
For the current epoch, we have set the scale factor to unity. Like earlier, we consider the background field solution ansatz to be
\begin{equation}
\Phi = \sigma_{0}e^{i\lambda\eta} \implies \lambda^{2} = m^{2} - \zeta \sigma_{0}^{2}
 + \xi\sigma_{0}^{4},
\end{equation}
where $\sigma_{0}$ is the vacuum expectation value of the field $|\Phi|$. Now, we consider the fluctuation around it, which is of the same form
\begin{equation}
\Phi = (\sigma_{0} + h)e^{i\lambda\eta + \pi},
\end{equation}
where $h$ is massive fluctuation mode and $\pi$ describes massless phonon modes \cite{berezhiani2015theory}. Plugging the above field ansatz into the Lagrangian density for 
Q-ball and consider the lowest order perturbations in $\pi$, we obtain the following Lagrangian 
density
\begin{equation}
\begin{split}
\mathcal{L}_{\text{pert}} & \sim (\sigma_{0} + h)^{2}[2\mu\dot{\pi} - (\vec{\nabla}\pi)^{2} + \xi 
\sigma_{0}^{4} - \zeta\sigma_{0}^{2}] + \frac{\zeta}{2}(\sigma_{0} + h)^{4}\\
 & - \frac{\xi}{3}(\sigma_{0} + h)^{6}.
\end{split}
\end{equation}
Now, integrating out the $h$ degrees of freedom leads to the following relation:
\begin{equation}
(\sigma_{0} + h)^{2} = \frac{\zeta + \sqrt{\zeta^{2} + 4\xi X}}{2\xi},
\end{equation} 
where $X$ is given by:
\begin{equation}
X = 2\lambda\dot{\pi} - (\vec{\nabla}\pi)^{2} + \xi\sigma_{0}^{4} - \zeta\sigma_{0}^{2}.
\end{equation}
As a result of the relation mentioned above, we obtain the following expression of $\mathcal{L}_{\text{pert}}$
\begin{equation}
\mathcal{L}_{\text{pert}} = \frac{\zeta^{3} + 12\zeta\xi X + (\zeta^{2} + 4\xi X)^{3/2}}
{12\xi^{2}}.
\end{equation}
We know from renormalization group flow \cite{Peskin:1995ev} of $\phi^{4}$-theory that
\begin{equation}
\zeta(p) = \frac{\zeta(m)}{1 - \frac{3\zeta(m)}{16\pi^{2}}\log\left(\frac{p}{m}\right)},
\end{equation}
which means in the IR limit (galaxy scale) $\zeta\rightarrow 0$ and this conclusion is valid even after the inclusion of hexic term in the field potential since its coupling $\xi$ (in our model) is irrelevant at the IR in four spacetime dimension. Therefore, for $\zeta \ll \xi m^{2}$ and for $X > 0$, the above Lagrangian density reduces to the following form
\begin{equation}
\mathcal{L}_{\text{pert}} \sim \frac{2}{3\sqrt{\xi}}X\sqrt{X},
\end{equation}
which has the same structure as the one conjectured in \cite{berezhiani2015theory} for explaining
MOND at the galaxy scale using the superfluid dark matter model (here by MOND we meant a scalar fluid with observational properties similar to the MOND paradigm). However, unlike in Ref.~\cite{berezhiani2015theory}, we do not start with a Lagrangian density containing a term like $|\partial\Phi|^{6}$. In Ref.~\cite{berezhiani2015theory}, the authors argue that MOND cannot be derived from a pure $|\Phi|^{6}$ potential, representing our limiting case. By examining the sign of $X$, we can prove that this conclusion is incorrect. Since we are considering phonons in the long-wavelength limit, the term $(\vec{\nabla}\pi)^{2}/2m$ is much smaller compared to the other three terms in $X$. Moreover, in the low-energy limit, the expectation value of $\sigma_{0}$ approaches zero, as shown in Eq. (\ref{eq. 38}). Therefore, our approximation aligns with the initial assumption, indicating that a MOND-like theory can emerge from the low-energy effective theory of Q-ball matter.

\subsection{Consequences of Q-ball DM superfluid}

The galaxy merger rate is one interesting consequence of the superfluid DM compared to CDM. Using Landau's criterion for superfluidity, two possible results can be obtained depending on the infall speed. If the infall velocity is less than the speed of sound, then the halos of the galaxy condensate will pass through each other with minimal dissipation. This will result in a much longer time scale for the merger than in the $\Lambda$CDM model and multiple encounters, as the density of superfluid halos is negligible. Conversely, suppose the infall velocity exceeds the sound speed. In that case, encounters will cause the halos to be out of equilibrium and attract DM particles from the condensate, resulting in a fast halo merger, similar to $\Lambda$CDM. After a while, the merged halo undergoes thermalization and condensates back to its superfluid ground state. This needs a detailed investigation, which we plan to do in the following work.

As we have shown above, MOND (here we considered scalar fluid with observational properties similar to the actual MOND) emerges out of the superfluid phase of Q-ball matter at the galactic scale~\cite{berezhiani2015theory}. This establishes the fact that Q-balls mimic CDM at cosmological scale whereas it behaves as collective excitation at galactic scale. Thus, it establishes the possible connection between CDM and MOND at two different length scales which are mutually exclusive.

\section{Discussion}\label{section Discussion}

As discussed earlier, the CDM model explains CMB temperature variations and structure formation on cosmic scales. However, the CDM has consistently been questioned by measurements that look at the deepest dark matter halo areas and the characteristics of dwarf galaxy satellites. In contrast, a wide variety of galactic events may consistently be explained by MOND. MOND, however, cannot account for the CMB and matter power spectra's intricate shapes. In order to resolve this contrast, we proposed millicharged composite Q-balls as a candidate for dark matter in the present article. 

In order to show this, we first established that composite Q-balls can indeed form in the RD epoch and drive it towards the MD epoch of the Universe. We showed that the interaction between the complex scalar field and the radiation is irrelevant in the FRW background. We then obtained two conditions for the Q-ball formation in the early Universe. 
Moreover, we have also shown that the formation of the Q-balls in the RD epoch is robust by considering the thermal corrections. Then, we computed the number density of Q-balls in the MD epoch and its formation rate, using which we could put an upper bound on the mass of the Q-balls. From the matter-radiation equality, we also obtain the mass of Q-balls to be $1~{\rm eV}$, which are much smaller than the electron mass. Combined with the invisible decay mode of ortho-positronium leads to $Q < 3.4 \times 10^{-5}$ \cite{PhysRevD.88.117701, PhysRevD.75.032004, PhysRevD.75.075014}. Suggesting that the millicharged Q-balls are DM candidates responsible for the early structure. Based on the experimental data, this consistency check bolsters the millicharged Q-balls with mass less than 1~eV as a possible dark matter candidate in the cosmological scales.

The Q-ball model outlined in this article differs significantly from axions due to the field potential's independence from the angular field variable $\theta$ of the complex scalar field. Instead, in the background geometry, this field variable assumes a non-zero value that is contingent upon the global charge $Q$ of the system. Furthermore, as detailed in section \ref{section 1}, it becomes apparent that the value $\lambda$, which is associated with the angular field variable $\theta$ up to the conformal time $\eta$, serves as a chemical potential of the system. This distinction effectively separates our model from axion field theories. Although the Q-ball model outlined in this article possesses all the properties of WIMPs, there are distinct differences between these two types of theories. Firstly, Q-balls are formed specifically during the radiation-dominated epoch through the solitosynthesis mechanism discussed earlier, whereas, for WIMPs, the formation channel is entirely different (see \cite{roszkowski2018wimp}). Secondly, Q-balls' formation rate and number density depend on the global $U(1)$ charge $Q$, distinguishing them from WIMPs. Thirdly, WIMPs are massive, whereas Q-balls shown here are lighter. Furthermore, WIMPs cannot form the superfluid phase of dark matter at the galactic scale, whereas Q-balls could do so through the formation of BEC. These features clearly distinguish Q-balls from other candidates of dark matter.   

In order to show that Q-balls can lead to MOND at the galactic scales, we looked at the possibility of the formation of BEC and superfluidity by Q-balls as suggested in \cite{khoury2022dark, berezhiani2015theory}, and we found that Q-balls can indeed form a BEC given the averaged matter density at the galactic scales. Moreover, it is also shown that once the Q-balls form through the charge accumulation process and form BEC, they can give rise to a superfluid state of matter due to a contact-repulsive interaction that arises due to the charge repulsion. We explicitly demonstrated in Sec.~\ref{MOND from DM superfluid} how MOND naturally emerges at galactic scales from the low-energy effective theory of phonons surrounding the proposed superfluid dark matter, as detailed in \cite{berezhiani2015theory, PhysRevD.95.043541, Berezhiani:2015pia}. In contrast to the approach outlined in \cite{berezhiani2015theory}, our analysis does not contain $|\partial\Phi|^{6}$ in the Lagrangian. Unlike in Ref.~\cite{berezhiani2015theory}, we have shown that MOND can be derived from a purely $|\Phi|^{6}$ potential. This demonstrates that the Q-balls can provide a unified framework for the dark matter at galactic and cosmological scales.

Thus Q-balls act differently in cosmological and galactic scales. In the cosmological scales, as Q-balls begin to clump together to form a halo, they behave similarly to CDM. Up to the point where particles start to scatter off each other, they can mimic the density profile of a typical galaxy. However, in regions where the density becomes high enough for phonons to frequently collide, the evolution of the density profile starts to diverge from that of a collisionless system. Thus, as shown explicitly, the Q-ball at the galactic scale which can mimic the MOND behaviour. As has been shown earlier, MOND can solve the core-cusp problem and the too-big-to-fail problem in galaxy formation. Thus, we expect that the superfluid state of the Q-ball in galactic scales can potentially solve these problems. In a follow-up work, we plan to identify possible signatures that could effectively distinguish Q-balls from dark matter observations.

In our future study, we want to look at the possibility of vortex formation by Q-balls at the galactic scales that were previously studied through different mechanisms \cite{hui2021vortices,mielczarek2010vortex,mauland2022quantized}. Lastly, from Eq.~(\ref{eq. 74}), we may write $\rho_{Q}$ in terms of the scale factor as 
\begin{equation}
\rho_{Q} =  \frac{\rho_{Q}^{(0)}}{a^{3/2}} \sim \rho_{Q}^{(0)} z^{3/2}.
\end{equation}
Substituting in the Friedmann equation, we infer that $H_{MD}(z) \sim H_{0}(1 + z)^{3/4}$ which is distinct from that of baryonic matter. The implications of this for the structure formation are beyond the scope of this work and require further investigation. We also plan to modify the publicly available N-body code with the Q-ball modified expansion and look at the structure formation in the non-linear regime.  

\section{Acknowledgement}
The authors thank A. Kushwaha for comments on the earlier draft. SM is supported by SERB-Core Research Grant (Project RD/0122-SERB000-044).
\appendix

\section{Interaction between perturbations around complex scalar field and U(1) gauge field}
\label{app:U1gauge}
In this section, we discuss how the Q-balls get decoupled from the gauge field in the background and the interaction are present only at the level of perturbations. Although this is known result, we show this explicitly for completeness~\cite{Kushwaha:2019rxj}. To see this, let us consider the following $U(1)$ gauge field theory in arbitrary 4-D space-time:
\begin{equation}
S = - \int\sqrt{-g}d^{4}x \Big[g^{\mu\nu}(\mathcal{D}_{\mu}\Phi)(\mathcal{D}_{\nu}\Phi)^{*} + U(\Phi\Phi^{*}) + \frac{1}{4}F_{\mu\nu}F^{\mu\nu}\Big],
\end{equation}
where $\mathcal{D}_{\mu} = \partial_{\mu} - iqA_{\mu}$ is the gauge covariant derivative, and $F_{\mu\nu} = \nabla_{\mu}A_{\nu} - \nabla_{\nu}A_{\mu} = \partial_{\mu}A_{\nu} - \partial_{\nu}A_{\mu}$ is the field-strength tensor associated with the gauge field. Substituting the expression of covariant derivative, the above action can be rewritten as
\begin{equation}
\begin{split}
S = - \int\sqrt{-g}d^{4}x \Big[ & g^{\mu\nu}\partial_{\mu}\Phi\partial_{\nu}\Phi^{*} + J_{\mu}A^{\mu} + q^{2}\Phi^{*}\Phi A_{\mu}A^{\mu}\\
 & + \frac{1}{4}F_{\mu\nu}F^{\mu\nu} + U(\Phi^{*}\Phi)\Big], 
\end{split}
\label{eq:U1action}
\end{equation}
where $J_{\mu} = iq(\Phi^{*}\partial_{\mu}\Phi - \partial_{\mu}\Phi^{*}\Phi)$ is the conserved $U(1)$ four-current. It is important to note that the above gauge field does not drive the radiation-dominated epoch. In the early universe, the perturbations can be assumed to be small. Hence, we can decompose the field variables as follows:
\begin{equation}
\begin{split}
\Phi(\eta,{\bf x}) & = \bar{\Phi}(\eta) + \delta\Phi(\eta,{\bf x}) \, , \Phi^{*}(\eta,{\bf x}) = \bar{\Phi}^{*}(\eta) + \delta\Phi(\eta,{\bf x})\\  
A_{\mu}(\eta,{\bf x}) & = \bar{A}_{\mu}(\eta) + \delta A_{\mu}(\eta,{\bf x})
\end{split}
\end{equation}
where $\bar{\Phi}(\eta)$ is the field configuration associated with the FRW background and,  similarly,  $\bar{A}_{\mu}$ is the background gauge field.  $\delta\Phi, \ \delta\Phi^{*}$ and $\delta A_{\mu}$ are the perturbations about the background. \\
The conductivity of the medium during most of the history of the universe is(was) very high. [During the radiation-dominated era, the universe was a plasma, and during/after the reionization era, again, the universe was a plasma.] Hence,  there are no electric fields. Although magnetic fields can be present,  these are of the order of $10^{-15}~{\rm Gauss}$ as the large value will break the background isotropy and is not seen in CMB~\cite{Turner:1987bw,Kushwaha:2021csq}. The Coulomb gauge condition implies ${A}_0 = 0$ and $\partial_i A^i = 0$. To satisfy the homogeneity and isotropy of the FRW background, the U(1) gauge field must satisfy the condition $\bar{A}_i =0$. With the above-mentioned decomposition, the action \eqref{eq:U1action} can be written as:
\begin{equation}
S = S_{0} + S_{1},
\end{equation}
where
\begin{equation}
S_{0} = - \int\sqrt{-g}d^{4}x \Big[g^{\mu\nu}\partial_{\mu}\bar{\Phi}\partial_{\nu}\bar{\Phi}^{*} + U(\bar{\Phi}^{*}\bar{\Phi})\Big]\,  .
\end{equation}
%
The above expression of the classical (tree-level) action $S_{0}$ shows that there is no interaction between Q-balls and photons. This is the same ansatz of the Q-ball field configuration mentioned earlier (cf. Eq. \eqref{eq. 4}), and the fact that we can always choose a gauge in which $A^{0} = 0$.    
$S_{1}$ is given by
\begin{equation}
\begin{split}
S_{1} & = - \int\sqrt{-g}d^{4}x \Big[\frac{1}{4} \delta F_{\mu\nu}\delta F^{\mu\nu} + q^{2}\bar{\Phi}^{*}\bar{\Phi} \left(\delta A_{\mu} \delta A^{\mu}\right)\\
& + \delta J_{\mu} \delta A^{\mu} + g^{\mu\nu}\partial_{\mu}\delta\Phi^{*}\partial_{\nu}\delta\Phi + \frac{1}{2}\frac{\partial^{2}U}{\partial\Phi^{2}}\Big|_{\bar{\Phi}}\delta\Phi^{2}\\ 
&  + \frac{1}{2}\frac{\partial^{2}U}{\partial\Phi^{*2}}\Big|_{\bar{\Phi}}\delta\Phi^{*2} + \frac{\partial^{2}U}{\partial\Phi\partial\Phi^{*}}\Big|_{\bar{\Phi}}\delta\Phi\delta\Phi^{*}\\
& + {q^{2}(\delta\Phi^{*}\bar{\Phi} + \bar{\Phi}^{*}\delta\Phi + \delta\Phi^{*}\delta\Phi) \delta A_{\mu}\, \delta A^{\mu}} \Big],
\end{split}
\label{eq:S1}
\end{equation}
where the fluctuation in current $\delta J_{\mu}$ is given by
\begin{equation}
\begin{split}
\delta J_{\mu} & = iq(\bar{\Phi}^{*}\partial_{\mu}\delta\Phi - \partial_{\mu}\bar{\Phi}^{*}\delta\Phi) + 
iq(\delta\Phi^{*}\partial_{\mu}\bar{\Phi} - \partial_{\mu}\delta\Phi^{*}\bar{\Phi})\\
 & + iq(\delta\Phi^{*}\partial_{\mu}\delta\Phi - \partial_{\mu}\delta\Phi^{*}\delta\Phi).
\end{split}
\end{equation}
From the above, we see that $S_{0}$ describes the classical action whose minimization subject to the Q-ball ansatz Eq. \eqref{eq. 15} leads to the radial profile of Q-ball configuration. $S_{1}$ describes the interactions between  the gauge field fluctuations and $\delta\Phi$. Hence, in the background FRW, there is no interaction between the gauge field and $\bar{\Phi}$; hence, action \eqref{eq. 4} describes the Q-ball configuration in the radiation-dominated epoch.
In other words, Q-balls are decoupled from the gauge field soon after the formation in the RD epoch.   \\
As seen from the last term of the $S_{1}$ action \eqref{eq:S1}, the interaction exists between the fluctuations of these two fields. However, the interaction is in the third and fourth order. Also, from the second term of the action $S_1$, the interaction between the gauge field and $\delta\Phi$ is in short range. This validates the non-trivial interaction between the complex scalar and the gauge field used in Eq. \eqref{eq. 59}.

\section{Thin-wall Q-balls}\label{app:thinwall}

In this section, we go through the thin-wall approximation for the Q-ball configuration in the radiation-dominated epoch since this approximation is valid for Q-balls with large sizes. Moreover, to transition from the RD epoch to a matter-dominated epoch, Q-ball configurations should be dominated over the radiation and the normal matter constituents of the Universe, which requires the charge to be high so that the energy density eventually dominates over the radiation. While the calculations in this section are similar to the flat space-time case~\cite{coleman1985q}, there are some differences that we want to highlight.

Thin-wall Q-ball essentially corresponds to the large volume limit in which the energy of the Q-ball is dominantly contributed by the homogeneous core of the Q-ball (due to the large charge). The thin wall interpolates the core value of the field $\sigma_{0}$ to the trivial vacuum $\sigma = 0$ of the theory. 
In this limit, the Q-ball profile can be approximated to the following form:
\begin{equation}\label{eq. 30}
\sigma(r) \approx \sigma_{0}\Theta(R - r),
\end{equation}
where $R$ is the radius of the Q-ball. We may note that the above ansatz is consistent with the boundary conditions of the Q-ball. As the core is homogeneous, we can effectively express the energy density of the Q-ball in the thin-wall approximation as 
\begin{equation}
\label{eq:thinwall-EQ}
E_{Q} = \lambda_{0}Q   
\end{equation}
where $\lambda_0$ is the Lagrange multiplier defined in Sec. \eqref{section 1} corresponding to the thin-wall Q-balls. This above expression is valid because the derivative terms contribute only over a thin region. 

Substituting the above form of $\sigma(r)$ in Eq.~(\ref{eq. 16}) reduces to:
\begin{equation}\label{eq. 31}
\mathcal{E}_{\lambda} \approx \lambda_{0} Q + V\left(- \frac{1}{2}\lambda_{0}^{2}\sigma_{0}^{2} + \bar{U}(\sigma_{0}) \right) \, .
\end{equation}
where $V = \frac{4\pi}{3}R_{Q}^{3}$ is the volume of the Q-ball. $\mathcal{E}_{\lambda}$ is a function of $\lambda_0, \sigma_0, Q, R_Q$ (hence, $V$). 
In order to determine $\lambda_{0}$, we need to minimize the above expression \textit{w.r.t} $\lambda_{0}, \sigma_0, Q, R_Q$. This will lead to four independent relations, leading to a unique solution. 

First, minimizing Eq.~\eqref{eq. 31} w.r.t $\lambda_0$ leads to:
\begin{equation}\label{eq. 32}
Q = \lambda_{0}\sigma_{0}^{2}V.
\end{equation}
Substituting the above condition in Eq.~\eqref{eq. 31} results in:
\begin{equation}\label{eq. 33}
\mathcal{E}_{\lambda} \approx \frac{Q^{2}}{2\sigma_{0}^{2}V} + \bar{U}(\sigma_{0})V.
\end{equation}
In order to determine the volume of the core, we minimize the above expression \textit{w.r.t} the volume. This leads to the following relation:
\begin{equation}\label{eq. 34}
V^{2} = \frac{Q^{2}}{2\sigma_{0}^{2}\bar{U}(\sigma_{0})}~{\rm or}~
V = \frac{Q}{\sigma_{0} \sqrt{2 \bar{U}(\sigma_{0})}}
\end{equation}
Substituting the above relation in Eq.~\eqref{eq. 33} leads to:
\begin{equation}\label{eq. 35}
\mathcal{E}_{\lambda} = 2V \bar{U}(\sigma_{0})
= Q\sqrt{\frac{2\bar{U}(\sigma_{0})}{\sigma_{0}^{2}}} \, .
\end{equation}
Comparing the above expression with Eq. \eqref{eq:thinwall-EQ}, gives the following relation:
\begin{equation}\label{eq. 36}
\lambda_{0} = \sqrt{\frac{2 \bar{U}(\sigma_{0})}{\sigma_{0}^{2}}}.
\end{equation}
However, to determine $\lambda_0$ uniquely, we need to determine $\sigma_0$. 
This can be achieved by minimizing $\mathcal{E}_\lambda$ \textit{w.r.t} $\sigma_{0}$ such that $\sigma_{0} < m_{\sigma}$
\begin{equation}\label{eq. 37}
\frac{d\bar{U}(\sigma_{0})}{d\sigma_{0}} = 2\frac{\bar{U}(\sigma_{0})}{\sigma_{0}}.
\end{equation}
For the Hexic potential \eqref{eq:Poten}, the above relation reduces to
\begin{equation}\label{eq. 38}
\frac{\zeta}{2}\sigma_{0}^{3} = \frac{2\xi}{3a^{2}}\sigma_{0}^{5} \implies \sigma_{0} = \sqrt{\frac{3a^{2}\zeta}{4\xi}}.
\end{equation}
Substituting the above expression in Eq.~\eqref{eq. 36} leads to the following form:
\begin{equation}\label{eq. 39}
\begin{split}
\lambda_{0} & = \sqrt{\frac{2 U(\sigma_{0})}{\sigma_{0}^{2}}} = a\sqrt{m_{\sigma}^{2} - \frac{3\zeta^{2}}{16\xi}} < m_{\sigma} a.
\end{split}
\end{equation}
Substituting the above expression in Eq.~\eqref{eq:thinwall-EQ}, we have:
\begin{equation}
E_Q = a(\eta) Q \sqrt{m_{\sigma}^{2} - \frac{3\zeta^{2}}{16\xi}}
\label{eq:thinwall-EQfin}
\end{equation}
From the above expressions, we infer the following:
First, substituting the above expression in Eq. \eqref{def:Tbar-Deltabar}, we have:
\begin{equation}
\bar{\Delta}^{2} = a^{2} - \bar{\lambda}_{0}^{2} = \frac{3\zeta^{2}a^{2}}{16\bar{\xi}} > 0 \, .   
\end{equation}
Comparing this with the condition \eqref{eq:QBall-Stab}, we see that the Q-balls produced during the RD era are stable. This justifies the thin-wall approximation and the formation of Q-balls in the RD epoch.
Second, the above expression shows that the core value of the field increases in the RD epoch. This implies that the energy of the Q-ball in the thin-wall approximation \eqref{eq. 30} increases justifying the thin-wall approximation at late-times. 
Third, from the expression \eqref{eq:thinwall-EQfin}, the energy density associated with the Q-ball in the RD epoch is
\begin{equation}
\rho_{Q} = (E_{Q}/a)n_{Q} \sim \frac{1}{a^{3}(\eta)}
\end{equation}
Note that the energy density of radiation $\rho_{\gamma}$ decays as $1/a^{4}$. This result is crucial for establishing that the Q-balls can act as dark matter and in cosmological scales can be treated as CDM. We discuss more on this in Sec. III. Lastly, using the condition that $\lambda_0 > 0$, we have: 
\[m_{\sigma} > \sqrt{\frac{3\zeta^{2}}{16\xi}}~~\Longrightarrow~~1 > \sqrt{\frac{3\zeta^{2}}{16\bar{\xi}}} \, . 
\]
Using Eq.~\eqref{eq. 34} and substituting Hexic potential \eqref{eq:Poten}, we have: 
\begin{equation}\label{eq. 40}
R_{Q} 
= \left(\frac{Q \, \xi}{\pi\zeta a^3 \sqrt{\frac{1}{2}m_{\sigma}^{2} - \frac{3\zeta^{2}}{32\xi}}}\right)^{1/3} \, .
\end{equation}
Thus, we see that $R_Q$ increases as $Q$ increases.

Here we want to emphasize a few points regarding the structure of Q-balls within thin-wall approximation. Firstly, from the equation \eqref{eq. 6}, one might think that the non-trivial vacuum expectation value inside the core of the Q-ball within the thin-wall approximation gets large due to the cosmic expansion.  However, that is not true. This can be seen by combining equation \eqref{eq. 38} obtained from the minimization of $\mathcal{E}_{\lambda}$ from the various parameters with the equation \eqref{eq. 6}. Moreover, from the expression \eqref{eq. 40} one might think the size of the Q-ball decreases with the cosmic expansion despite it being a local object. However, that is not true if we express in-terms of the comoving-radius:
\begin{equation}\label{Co-moving radius of Q-balls}
aR_{Q} = \left(\frac{Q \, \xi}{\pi\zeta\sqrt{\frac{1}{2}m_{\sigma}^{2} - \frac{3\zeta^{2}}{32\xi}}}\right)^{1/3}.
\end{equation}
This shows that the Q-ball configuration and its size are not affected by the cosmic expansion.

\section{Classical Stability of Q-balls in RD epoch}
\label{app:Derricktheorem}
Following Ref. \cite{PaccettiCorreia:2001wtt}, the condition for the classical stability of Q-balls is:
\begin{equation}
\frac{\lambda}{Q}\frac{dQ}{d\lambda} < 0.
\end{equation}
As we have seen earlier, the existence of Q-ball demands the following condition on $\lambda^{2}$
\begin{equation}
\lambda_{0}^{2} < \lambda^{2} < m_{\sigma}^{2}a^{2}, \lambda_{0}^{2} = \text{min}\left(\frac{2\bar{
U}(\sigma)}{\sigma^{2}}\right) \equiv \frac{2\bar{U}(\sigma_{0})}{\sigma_{0}^{2}},
\end{equation}
where $\sigma_{0}$ is the minima of $2\bar{U}(\sigma)/\sigma^{2}$. In this work, we have considered $\lambda_{0}^{2} > 0$, which, of course, restricts the parameter space of field potential. [See, for instance, Eq. \eqref{existence condition 2}.] Thus, $\lambda$ is considered positive, and consequently, $Q$ is positive. For thin-wall Q-balls, as discussed in Appendix \ref{app:thinwall}, we have 
\begin{equation}
\frac{dQ}{d\lambda} = - 
\frac{5 \lambda^{2} + \lambda_0^2}{\lambda^{2} - \lambda_{0}^{2}} < 0.
\end{equation}
The above result clearly shows that the Q-balls within thin-wall approximation are classically stable. Since the above inequality follows the existence criterion of Q-balls in the RD epoch, it holds independent of the scale factor $a$ value in the RD epoch.
%
%
%
%
%
%
%

\section{Bose-Einstein Condensation}\label{app:BEC}
Here, we briefly discuss some facts about Bose-Einstein condensation (BEC), as some readers may not be familiar with the mathematical details of BEC (see also \cite{khoury2022dark}). The partition function in the grand canonical ensemble of charged particles with chemical
potential $\mu$ and single particle dispersion relation $\epsilon_{\vec{k}}$ is given by
\begin{equation}\label{eq. 2.35}
\mathcal{Z} = \sum_{\vec{k}}\text{Tr}\Big[e^{\beta\hat{N}_{\vec{k}}(\mu - \epsilon_{\vec{k}})}
\Big],
\end{equation} 
where $\hat{N}_{k}$ is the number operator associated with the mode labeled by $\vec{k}$. The average occupation number in such a state is given by $\langle N_{\vec{k}} \rangle = \sum_{N_{\vec{k}} = 0}^{\infty}N_{\vec{k}} \, P(N_{\vec{k}})$ where the probability
density is given by
\begin{equation}\label{eq. 2.36}
P(N_{\vec{k}}) = \frac{e^{\beta N_{\vec{k}}(\mu - \epsilon_{\vec{k}})}}{\mathcal{Z}}.
\end{equation}
The above sum converges and leads to the following Bose-Einstein distribution of the number of particles in $\vec{k}$ mode:
\begin{equation}\label{eq. 2.37}
\langle \hat{N}_{\vec{k}}\rangle = \frac{1}{e^{\beta(\epsilon_{\vec{k}} - \mu)}},
\end{equation}
if $\mu < \epsilon_{\vec{k}}$ for all $\vec{k}$. Summing over all the $\vec{k}$ modes gives the total number of particles:
\begin{equation}\label{eq. 2.38}
\begin{split}
N & = \sum_{\vec{k}}\langle\hat{N}_{\vec{k}}\rangle = N_{0} + N_{exc}\\
 & = \frac{1}{e^{-\beta\mu} - 1} + \sum_{\vec{k}\neq 0}\frac{1}{e^{\beta(\epsilon_{\vec{k}} - \mu)} - 1},
\end{split}
\end{equation}
where $N_{0}$ is the ground state occupancy, assuming $\epsilon_{0} = 0$ without loss of generality and $N_{exc}$ is occupation number for excited states. In terms of fugacity, defined by $z = e^{\beta\mu}$, we may express $N_{0}$ as
\begin{equation}\label{eq. 2.39}
N_{0} = \frac{z}{1 - z}.
\end{equation}
In the case of the ground state, the convergence of the geometric series in (\ref{eq. 2.38}) requires $- \infty < \mu < 0$, which implies $0 < z < 1$. 

In the MD epoch, the particles are non-relativistic and we can consider non-relativistic parabolic dispersion $\epsilon_{\vec{k}} = \vec{k}^{2}/{(2m)}$. Assuming that the gravitational field is weak and that the Q-balls formed during the RD epoch weakly self-interact, we may express $N_{exc}$ as 
\begin{equation}\label{eq. 2.40}
\begin{split}
N_{exc} & = V\int\frac{d^{3}k}{(2\pi)^{3}}\frac{1}{e^{\beta(\epsilon_{\vec{k}} - \mu)} - 1}\\
 & = V\frac{m^{3/2}}{\sqrt{2}\pi^{2}}\int_{0}^{\infty}d\epsilon \frac{\epsilon^{1/2}}{e^{\beta
(\epsilon - \mu)} - 1},
\end{split}
\end{equation}
where $V$ is the volume of the system and the above integral leads to the following relation
\begin{equation}\label{eq. 2.41}
N_{exc} = V\frac{m^{3/2}}{\sqrt{2}\pi^{2}\beta^{3/2}}\int_{0}^{\infty}dx \frac{x^{1/2}}{z^{-1}e^{x} - 1} = \frac{V}{\lambda_{th}^{3}}g_{3/2}(z),
\end{equation}
where $\lambda_{th} = \sqrt{\frac{2\pi}{m k_{B}T}}$ is the thermal de Broglie wavelength, and $g_{\nu}(z)$ is the polylogarithm function defined by
\begin{equation}\label{eq. 2.42}
g_{\nu}(z) = \frac{1}{\Gamma(\nu)}\int_{0}^{\infty}dx \frac{x^{\nu - 1}}{z^{-1}e^{x} - 1} = 
\sum_{n = 1}^{\infty}\frac{z^{n}}{n^{\nu}}.
\end{equation}
Since the $g_{3/2}(z)$ is a monotonic increasing function and $z < 1$, we may say that $g_{3/2}(z) < \zeta(3/2)$, hence the number density of the excited state is bounded from above:
\begin{equation}\label{eq. 2.43}
n_{exc} < \frac{\zeta(3/2)}{\lambda_{th}^{3}}.
\end{equation} 
At a fixed temperature $T$, if we increase the number density by adding more particles such that
\begin{equation}\label{eq. 2.44}
n > \frac{\zeta(3/2)}{\lambda_{th}^{3}},
\end{equation}
then excess particles must populate the ground state, and this macroscopic occupation of the ground state is known as the BEC phenomenon. At fixed $T$, it occurs at the critical density
\begin{equation}\label{eq. 2.45}
n_{c} = \frac{\zeta(3/2)}{\lambda_{th}^{3}} = \zeta(3/2)\left(\frac{mk_{B}T}{2\pi}\right)^{3/2},
\end{equation}
whereas at fixed number density, the condensation phenomenon occurs at the critical temperature
\begin{equation}\label{eq. 2.46}
T_{c} = \frac{2\pi}{mk_{B}}\left(\frac{n}{\zeta(3/2)}\right)^{2/3}.
\end{equation}
Combining the above results, one can show that below the critical temperature, the ground state occupancy satisfies the following relation:
\begin{equation}\label{eq. 2.47}
\frac{N_{0}}{N} = 1 - \left(\frac{T}{T_{c}}\right)^{3/2}.
\end{equation}
We may note that the criterion of BEC roughly translates from (\ref{eq. 2.44}) (ignoring numerical factors) to
\begin{equation}\label{eq. 2.48}
\lambda_{th} \geq n^{-1/3} \equiv l,
\end{equation}
where $l$ is the characteristic inter-particle separation. This implies that for BEC to occur, the particles' thermal de Broglie wavelength must be larger than the inter-particle separation.

\bibliographystyle{apsrev4-2}
\bibliography{Ver-final}

\end{document}